\documentclass[preprint,12pt,3p]{elsarticle}

\usepackage{amsmath,amssymb}
\usepackage[colorlinks,bookmarksopen,bookmarksnumbered,citecolor=red,urlcolor=red]{hyperref}
\usepackage{moreverb,dblfloatfix}
\usepackage{algorithmic}
\usepackage{algorithm}
\usepackage{graphicx}
\usepackage{xcolor}
\usepackage{upgreek}
\usepackage{subfigure}
\usepackage{rotating}
\usepackage{multirow,xspace}
\newcommand{\x}{\mathbf{x}}
\newcommand{\xa}{\mathbf{x}^\textnormal{a}}
\newcommand{\xb}{\mathbf{x}^\textnormal{b}}

\newcommand{\xk}{\mathbf{x}_k}

\newcommand{\y}{\mathbf{y}}
\newcommand{\yk}{\mathbf{y}_k}

\newcommand{\Nobs}{\textnormal{m}}
\newcommand{\nens}{\textsc{n}_\textnormal{ens}}
\newcommand{\nvar}{\textsc{n}_\textnormal{var}}
\newcommand{\nc}{\textsc{n}_\textnormal{c}}
\newcommand{\p}{\mathbf{p}}

\newcommand{\PD}{\mathcal{P}}
\newcommand{\Pa}{\mathcal{P}^{\rm a}}
\newcommand{\Pb}{\mathcal{P}^{\rm b}}
\newcommand{\muki}{\mathbf{\mu}_{k,i}}

\newcommand{\tauki}{\mathbf{\tau}_{k,i}}

\newcommand{\Upsigmai}{\mathbf{\Upsigma}_{i}}
\newcommand{\Upsigmaki}{\mathbf{\Upsigma}_{k,i}}
\newcommand{\ClHMC}{$\mathcal{C}\mathrm{\ell}\textnormal{HMC}$\xspace}

\journal{Tellus A}
\begin{document}
%======================================================================================================================

  %_______________________________
  % Include the Technical report coverpage
  \thispagestyle{empty}
\setcounter{page}{0}

\begin{Huge}
\begin{center}
Computer Science Technical Report CSTR-{4/2016} \\
\today
\end{center}
\end{Huge}
\vfil
\begin{huge}
\begin{center}
{\tt Ahmed Attia, Azam Moosavi, and Adrian Sandu}
\end{center}
\end{huge}

\vfil
\begin{huge}
\begin{it}
\begin{center}
``{\tt Cluster Sampling Filters for Non-Gaussian Data Assimilation}''
\end{center}
\end{it}
\end{huge}
\vfil

\begin{large}
\begin{center}
Computational Science Laboratory \\
Computer Science Department \\
Virginia Polytechnic Institute and State University \\
Blacksburg, VA 24060 \\
Phone: (540)-231-2193 \\
Fax: (540)-231-6075 \\ 
Email: \url{attia@vt.edu}, \url{azmosavi@vt.edu}, \url{sandu@cs.vt.edu} \\
Web: \url{http://csl.cs.vt.edu}
\end{center}
\end{large}

\vspace*{1cm}

\begin{tabular}{ccc}
\includegraphics[width=2.5in]{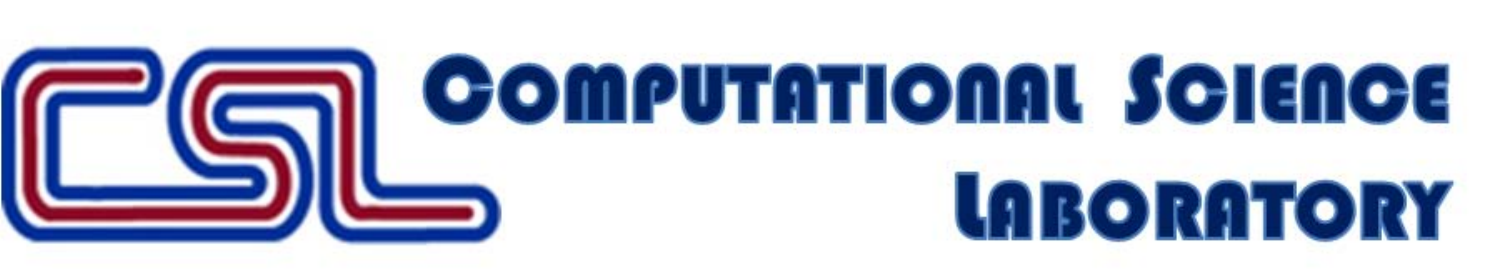}
&\hspace{2.5in}&
\includegraphics[width=2.5in]{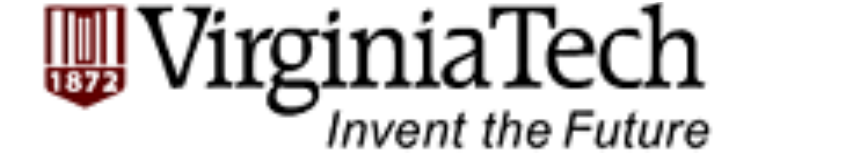} \\
{\bf\em Compute the Future!} &&\\
\end{tabular}

\newpage

  %_______________________________

  %_______________________________
  % Frontmatter
  \begin{frontmatter}

  %_______________________________
  %
  \title{Cluster Sampling Filters for Non-Gaussian Data Assimilation}
  \author[labela]{Ahmed Attia}
  \author[labela]{Azam Moosavi}
  \author[labela]{Adrian Sandu}
  \address[labela]{Computational Science Laboratory \\
  Department of Computer Science   \\
  Virginia Polytechnic Institute and State University \\
  2201 Knowledgeworks II, 2202 Kraft Drive, Blacksburg, VA 24060, USA \\
  Phone: 540-231-2193, Fax: 540-231-9218    \\
  E-mail: sandu@cs.vt.edu
  }
  %_______________________________
  %
  %
\begin{abstract}
This paper presents a fully non-Gaussian version of the Hamiltonian Monte Carlo (HMC) sampling filter.
The Gaussian prior assumption in the original HMC filter is relaxed. Specifically, a clustering step is introduced after the forecast phase of the filter, and the prior density function is estimated by fitting a Gaussian Mixture Model (GMM) to the prior ensemble.
Using the data likelihood function, the posterior density is then formulated as a mixture density, and is sampled using a HMC approach (or any other scheme capable of sampling multimodal densities in high-dimensional subspaces).
The main filter developed herein is named {\it cluster HMC sampling filter} (\ClHMC). A multi-chain version of the \ClHMC filter, namely MC-\ClHMC is also proposed to guarantee that samples are taken from the vicinities of all probability modes of the formulated posterior.
The new methodologies are tested using a quasi-geostrophic (QG) model with double-gyre wind forcing and bi-harmonic friction.
Numerical results demonstrate the usefulness of using GMMs to relax the Gaussian prior assumption in the HMC filtering paradigm.
\end{abstract}


\end{frontmatter}
%_______________________________

  \newpage
  \tableofcontents
  \newpage
  \setcounter{page}{1}
  %

%_______________________________________________________________________
  \section{Introduction} \label{Sec:Introduction}
%_______________________________________________________________________
Data assimilation (DA) is a complex process that involves combining information from different sources in order to produce accurate estimates of the true state of a physical system 
such as the atmosphere. Sources of information include computational models of the system, a background probability distribution, and observations collected at discrete time instances. 
With model state denoted by $\x\, \in \mathbb{R}^{\nvar}\,,$ the prior probability density $\Pb(\x)$ encapsulates the knowledge about the system state before incorporating any other source 
of information such as the observations.  
Let $\y\,\in \mathbb{R}^{\Nobs}$ be a measurement (observation) vector. The observation likelihood function $\PD(\y|\x)$ quantifies the mismatch between the model predictions 
(of observed quantities) and the collected measurements. 
A standard application of Bayes' theorem provides the posterior probability distribution $\PD(\x|\y)$ that provides an improved description of the unknown true state of the system of interest.

In the ideal case where the underlying probability distributions are Gaussian, the model dynamics is linear, and the observations are linearly related to the model state, 
the posterior can be obtained analytically for example by applying Kalman filter (KF) equations ~\cite{Kalman_1960,Kalman_1961}. 
For large dimensional problems the computational cost of the standard Kalman filter is prohibitive, and in practice the probability distributions are approximated using small ensembles. 
The ensemble-based approximation has led to the ensemble Kalman filter (EnKF) family of methods~\cite{Burgers_1998_EnKF,Evensen_1994,Evensen_2003,Houtekamer_1998a}. 
Several modifications of EnKF, for example~\cite{Hamill_2001,Houtekamer_2001,Whitaker_2002a,sakov2012iterative,smith2007cluster,Tippett_2003_EnSRF}, have been introduced in the literature 
to solve practical DA problems of different complexities. 

One of the drawbacks of the EnKF family is the reliance on an ensemble update formula that comes from the linear Gaussian theory. 
Several approaches have been proposed in the literature to alleviate the limitations of the Gaussian assumptions.
The maximum likelihood ensemble filter (MLEF)~\cite{lorenc1986analysis,zupanski2005maximum,zupanski2008maximum} computes the maximum a posteriori estimate of the state in the ensemble space. 
The iterative EnKF \cite{gu2007iterative,sakov2012iterative} (IEnKF) extends MLEF to handle nonlinearity in models as well as in observations. 
IEnKF, however, assumes that the underlying probability distributions are Gaussian and the analysis state is best estimated by the posterior mode.

These families of filters can generally be tuned (e.g., using inflation and localization) for optimal performance on the problem at hand. 
However, if the posterior is a multimodal distribution, these filters are expected to diverge, or at best capture a single probability mode, especially in the case of long-term forecasts. 
Only a small number of filtering methodologies designed to work in the presence of highly non-Gaussian errors are available, and their efficiency with realistic models is yet to be established.

The Hybrid/Hamiltonian Monte Carlo (HMC) sampling filter was proposed in~\cite{attia2015hmcfilter} as a fully non-Gaussian filtering algorithm, and has been extended to the four-dimensional 
(smoothing) setting in~\cite{attia2015hmcfilter,attia2015hmcsampling,attia2015hmcsmoother,attia2016reducedhmcsmoother}. 
The HMC sampling filter is a sequential DA filtering scheme that works by directly sampling the posterior probability distribution via an HMC approach~\cite{duane1987hybrid,toral1994general}. 
The HMC filter is designed to handle cases where the underlying probability distributions are non-Gaussian. 
Nevertheless, the first HMC formulation presented in~\cite{attia2015hmcfilter} assumes that the prior distribution can always be approximated by a Gaussian distribution. 
This assumption was introduced for simplicity of implementation;  however, it can be too restrictive in many cases, and may lead to inaccurate conclusions.  
This strong assumption needs to be relaxed in order to accurately sample from the true posterior, while preserving computational efficiency.

This work relaxes the Gaussian prior assumption in the original HMC formulation. Specifically, the prior is represented by a Gaussian Mixture Model (GMM) that is fitted to the forecast ensemble via a clustering step. 
The posterior is formulated accordingly. In the analysis step the resulting mixture posterior is sampled following a HMC approach. 
The analysis phase, however, can be easily modified to incorporate any other efficient direct sampling method. The resulting algorithm is named the cluster HMC (\ClHMC) sampling filter. 
In order to improve the sampling from the mixture posterior a more efficient version, named  \ClHMC filter (MC-\ClHMC), is also discussed.

Using a GMM to approximate the prior density, given the forecast ensemble, was presented in~\cite{anderson1999monte,smith2007cluster} as a means to solve the nonlinear filtering problem. 
In~\cite{anderson1999monte}, a continuous approximation of the prior density was built as a sum of Gaussian kernels, where the number of kernels is equal to the ensemble size. 
Assuming a Gaussian likelihood function, the posterior was formulated as a GMM with updated mixture parameters. 
The updated means and covariance matrices of the GMM posterior were obtained by applying the convolution rule of Gaussians to the prior mixture components and the likelihood, 
and the analysis ensemble was generated by direct sampling from  the GMM posterior. 
On the other hand, the approach presented in~\cite{smith2007cluster} works by fitting a GMM to the prior ensemble with number of mixture components detected using Akaike information criterion. 
The EnKF equations are applied to each of the components in the mixture distribution to generate an analysis ensemble from the GMM posterior.
    
Unlike the existing approaches \cite{anderson1999monte,smith2007cluster}, the methodology proposed herein is fully non-Gaussian, and does not limit the posterior density to 
a Gaussian mixture distribution or Gaussian likelihood functions.  
Here we sample the posterior distribution using a Hamiltonian Monte Carlo approach, however the direct sampling method can be replaced with any efficient analysis algorithm capable of 
dealing with high-dimensional multimodal distributions.

The remaining part of the paper is organized as follows.
Section~\ref{Sec:DA_and_HMC} reviews the original formulation of the HMC sampling filter. 
Section~\ref{Sec:GMM_and_ClHMC} presents the new algorithms \ClHMC and MC-\ClHMC.
Numerical results and discussions are presented in Section~\ref{Sec:Numerical_Results}.
Conclusions are drawn in Section~\ref{Sec:Conclusions}.
     
%____________________________________________________________________________
 \section{The HMC Sampling Filter} \label{Sec:DA_and_HMC}
 %____________________________________________________________________________
    %
    In this section we present a brief overview of the HMC sampling methodology, followed by the original formulation of the HMC sampling filter.

%~~~~~~~~~~~~~~~~~~~~~~~~~~~~~~~~~~~~~~~~~~~~~~~~~~~~~~~~~~~~~~~~~~~  % Probably we don't need the subsection header here!
\subsection{HMC sampling} \label{Subsec:HMC_sampling}
%~~~~~~~~~~~~~~~~~~~~~~~~~~~~~~~~~~~~~~~~~~~~~~~~~~~~~~~~~~~~~~~~~~~
%
HMC sampling follows an auxiliary-variable approach~\cite{besag1993spatial,sokal1997monte} to accelerate the sampling process of 
Markov chain Monte-Carlo (MCMC) algorithms. 
In this approach, the MCMC sampler is devoted to sampling the joint probability density of the target variable, 
along with an auxiliary variable. The auxiliary variable is chosen carefully to allow for the construction of a Markov chain that mixes faster, and is
easier to simulate than sampling the marginal density of the target variable~\cite{higdon1998auxiliary}. 
% It is also necessary that the marginalization of the density functions of the target variable and the auxiliary variable is easy to carry out.

The main component of the HMC sampling scheme is an auxiliary Hamiltonian system that plays the role of the proposal (jumping) distribution. %, also known as the jumping distribution.
The Hamiltonian dynamical system operates in a phase space of points  $(\p,\x) \in \mathbb{R}^{2\nvar}\,,$ where the individual
variables are the position $\x \in \mathbb{R}^{\nvar}\,,$ and the momentum $\p \in \mathbb{R}^{\nvar}$. 
The total energy of the system, given the position and the momentum, is described by the Hamiltonian function $H(\p,\x)$.
A general formulation of the Hamiltonian function (the Hamiltonian) of the system is given by:
\begin{equation} \label{eqn:Hamiltonian_function}
  H(\p,\x)  = \frac{1}{2} \, \p^T \mathbf{M}^{-1} \p - \log(\phi(\x)) = \frac{1}{2} \, \p^T \mathbf{M}^{-1} \p + \mathcal{J}(\x)\,,
\end{equation}  
where $\mathbf{M} \in \mathbb{R}^{\nvar\times\nvar}$ is a symmetric positive definite matrix referred to as the \textit{mass matrix}. The first term in the sum~\eqref{eqn:Hamiltonian_function} 
quantifies the kinetic energy of the Hamiltonian system, while the second term is the associated potential energy.

The dynamics of the Hamiltonian system in time is described by the following ordinary differential equations (ODEs):
\begin{eqnarray} \label{eqn:hamiltonian_equations}
    \frac{d\x}{dt} = \nabla_\p\, H\,,\qquad
    \frac{d\p}{dt} = - \nabla_\x\, H.
\end{eqnarray}
The time evolution of the system \eqref{eqn:hamiltonian_equations} in the phase space is described by the flow:~\cite{neal2011mcmc,sanz1994numerical}
\begin{equation} \label{eqn:hamiltonian_flow}
    \Phi_T:\mathbb{R}^{2\nvar} \rightarrow \mathbb{R}^{2\nvar}, \quad \Phi_T\bigl(\p(0),\x(0)\bigr)=\bigl(\p(T),\x(T)\bigr),
\end{equation}
which maps the initial state of the system $(\p(0),\x(0))$ to $(\p(T),\x(T))\,,$ the state of the system at time $T$.
In practical applications, the analytic flow $\Phi_T$ is replaced by a numerical solution using a time reversible and symplectic numerical integration 
method~\cite{sanz2014Markov,sanz1994numerical}. 
The length of the Hamiltonian trajectory $T$ can generally be long, and may lead to instability of the time integrator if the step size is set to $T$. 
In order to accurately approximate $\Phi_T\,,$ the symplectic integrator typically takes $m$ steps of size $h=T/m$ where $h$ is chosen such as to maintain stability. 
We will use $\Phi_T$ hereafter to represent the numerical approximation of the Hamiltonian flow.

Given the formulation of the Hamiltonian~\eqref{eqn:Hamiltonian_function}, the dynamics of the Hamiltonian system is governed by the equations
\begin{eqnarray} \label{eqn:hamiltonian_vector_dynamics}
\frac{d\x}{dt} =  \mathbf{M}^{-1} \p \,, \qquad
\frac{d\p}{dt} = -\nabla_\x \mathcal{J}(\x)\,. 
\end{eqnarray}

The canonical probability distribution of the state $(\p,\x)$ of the Hamiltonian system in the phase space $\mathbb{R}^{2\nvar}\,,$ upto a scaling factor, is given by
\begin{eqnarray} \label{eqn:Canonical_Pdf}
\exp{( - H(\p,\x) )}  =  \exp{\left( -\frac{1}{2} \p^T \mathbf{M}^{-1} \p - \mathcal{J}(\x) \right)} 
\propto \exp{\left( -\frac{1}{2} \p^T \mathbf{M}^{-1} \p \right)} \, \phi(\x). 
\end{eqnarray}
The product form of this joint probability distribution shows that the two variables $\p,\,\text{and } \x$ are independent~\cite{sanz2014Markov}. 
The marginal distribution of the momentum variable is Gaussian, $\p \sim \mathcal{N}(0,\mathbf{M})\,,$ 
while the marginal distribution of the position variable is proportional to the negative-logarithm (negative-log) of the potential energy, that is 
$\x \sim f(\x) \propto \phi(\x) = \exp{\left(-\mathcal{J}(\x)\right)}$.
Here $f(\x)$ is the normalized marginal density of the position variable, while $\phi(\x)$ drops the scaling factor (e.g. the normalization constant) of the density function.

In order to draw samples $\{\x(e)\}_{e=1,2,\ldots,\nens}$ from a given probability distribution $f(\x) \propto \phi(\x)\,,$ HMC makes the following analogy with 
the Hamiltonian mechanical system~\eqref{eqn:hamiltonian_equations}. 
The state $\x$ is \textit{viewed} as a position variable,  and an auxiliary momentum variable $\p\sim\mathcal{N}(0,\,\mathbf{M})$ is included.
The negative-log of the target probability density $\mathcal{J}(\x)=-\log(\phi(\x))$ is viewed as the potential energy of an auxiliary Hamiltonian system. 
The kinetic energy of the system is given by the negative-log of the Gaussian distribution of the auxiliary momentum variable. 
The mass matrix $\mathbf{M}$ is a user-defined parameter that is assumed to be symmetric positive definite. To achieve favorable performance of the HMC sampler,
$\mathbf{M}$ is generally assumed to be diagonal, with values on the diagonal chosen to reflect the scale of the components of the target variable under the target 
density~\cite{attia2015hmcfilter,neal2011mcmc}.
The HMC sampler proceeds by constructing a Markov chain whose stationary distribution is set to the canonical joint density~\eqref{eqn:Canonical_Pdf}.
The chain is initialized to some position and momentum values, and at each step of the chain, a Hamiltonian trajectory starting at the current state is constructed 
to propose a new state. A Metropolis-Hastings-like acceptance rule is used to either accept or reject the proposed state. 
Since both position and momentum are statistically independent, the retained position samples are actually sampled from the target density $ f(\x)$. 
The collected momentum samples are discarded, and the position samples are returned as the samples from the target probability distribution $ f(\x)$.

The performance of the HMC sampling scheme is greatly influenced by the settings of the Hamiltonian trajectory, that is the choice of the two parameters $m,\, h$.
The step size $h$ should be small enough to maintain stability, while $m$ should be generally large for the sampler to reach distant points in the state space.
The parameters of the Hamiltonian trajectory can be set empirically~\cite{neal2011mcmc} to achieve an acceptable rejection rate of at most $25\%~30\%\,,$ 
or be automatically adapted using automatic tuning schemes such as the No-U-Turn sampler(NUTS) \cite{hoffman2014NUTS}, 
or the Riemannian Manifold HMC sampler (RMHMC)~\cite{girolami2011riemann}.

The ideas presented in this work can be easily extended to incorporate any of the HMC sampling algorithms with automatically tuned parameters.
In this paper we tune the parameters of the Hamiltonian trajectory following the empirical approach, and focus on the sampler performance due to the choice of the prior distribution in the sequential filtering context. 

%~~~~~~~~~~~~~~~~~~~~~~~~~~~~~~~~~~~~~~~~~~~~~~~~~~~~~~~~~~~~~~~~~~~  % Probably we don't need the subsection header here!
\subsection{HMC sampling filter} \label{Subsec:HMC_sampling_filter}
%~~~~~~~~~~~~~~~~~~~~~~~~~~~~~~~~~~~~~~~~~~~~~~~~~~~~~~~~~~~~~~~~~~~
%
%Filtering refers to the assimilation of a single observation to the model forecast at a time. 
% The statistical solution of the filtering problem is an estimate of the true probability distribution of the unknown system state, given the prior.
%
In the filtering framework, following a perfect-model approach, the posterior distribution $\Pa(\xk)$ at a time instance $t_k$ follows from Bayes' theorem:
\begin{equation} \label{eqn:Bayes_Filtering_Rule}
\Pa(\xk) = \PD(\xk|\yk) 
 = \frac{\PD(\yk|\xk) \Pb(\xk) }{\PD(\yk) } \propto \PD(\yk|\xk) \Pb(\xk) \,,
\end{equation}
where $\Pb(\xk)$ is the prior distribution, $\PD(\yk|\xk)$ is the likelihood function, all at time instance $t_k$.
$\PD(\yk)$ acts as a scaling factor and is ignored in the HMC context.
% Here we follow a perfect-model approach.
% In the ensemble-based approach of data assimilation, the probability distributions involved in~\eqref{eqn:Bayes_Filtering_Rule} are estimated using a finite ensemble of states. 

As mentioned in Section~\ref{Sec:Introduction}, the formulation of the HMC sampling filter proposed in~\cite{attia2015hmcfilter} assumes that the prior distribution $\Pb(\xk)$ 
can be represented by a Gaussian distribution $\mathcal{N}(\xb_k,\, \mathbf{\mathbf{B}_k})\,,$ that is
\begin{equation} \label{eqn:Gaussian_Prior}
  \Pb(\xk) = \frac{(2\pi)^{-\frac{\nvar}{2}}}{\sqrt{|\mathbf{B}_k|}}\, 
\exp{\left( - \frac{1}{2}  \lVert \xk-\xb_k \rVert^2_ {\mathbf{B}_k^{-1}}  \right) }\,,
\end{equation}
where $\xb_k\,,$ is the background state, and $\mathbf{B}_k \in \mathbb{R}^{\nvar\times\nvar}$ is the background error covariance matrix.
The background state $\xb_k$ is generally taken as the mean of an ensemble of forecasts $\{\xb_k(e)\}_{e=1,\,2,\,\ldots,\,\nens}\,,$ obtained by forward model 
runs from a previous assimilation cycle. 
% Here the dimension of the system state space is $\nvar\,,$ and the ensemble size is $\nens$.
The associated weighted norm is defined as:
\begin{equation} \label{eqn:weighted_norm}
\mathbf{ \lVert a - b \rVert^2_ {\mathbf{C}} =  { (a - b)^T } {C} { (a - b) } }.
\end{equation}
Under the traditional, yet non-restrictive assumption, that the observation errors are distributed according to a Gaussian distribution with zero mean, and 
observation error covariance matrix $\mathbf{R}_k \in \mathbb{R}^{\Nobs\times\Nobs} \,,$ the likelihood function takes the form
\begin{equation} \label{eqn:GF_Likelihood}
  \PD(\yk|\xk) =\frac{(2\pi)^{-\frac{\Nobs}{2}}}{\sqrt{|\mathbf{R}_k|}}\, 
\exp{\left( - \frac{1}{2} \lVert \y_k - \mathcal{H}_k(\x_k) \rVert^2 _{ \mathbf{R}_k^{-1} }\right) }\,,
\end{equation}
where $\mathcal{H}_k : \mathbb{R}^{\nvar} \to \mathbb{R}^{\Nobs}$ is the observation operator that maps a given state $\x_k$ to the
observation space at time instance $t_k$.
The dimension of the observation space $\Nobs$ is generally much smaller than the state space dimension, that is $\Nobs \ll \nvar$.

The posterior follows immediately from~\eqref{eqn:Bayes_Filtering_Rule},~\eqref{eqn:Gaussian_Prior}, and~\eqref{eqn:GF_Likelihood} as:
\begin{subequations} \label{eqn:GF_Posterior}  % GF stands for Gaussian Framework
\begin{eqnarray}
\mathcal{P}^{\rm a}(\xk)  &\propto& \phi(\xk) \, = \, \exp{\Bigl( - \mathcal{J}(\xk) \Bigr)} \,, \label{eqn:GF_Potential_Energy} \\
\mathcal{J}(\xk) &=&   \frac{1}{2} \lVert \xk - \xb_k \rVert ^2 _ {\mathbf{B}_k^{-1} } 
     + \frac{1}{2}  \lVert \yk - \mathcal{H}_k(\xk)  \rVert ^2 _{ \mathbf{R}_k^{-1} }  \label{eqn:3DVar_cost_functional} \,, 
\end{eqnarray}
\end{subequations}
where $\mathcal{J}(\xk)$ is the negative-log of the posterior distribution~\eqref{eqn:GF_Potential_Energy}.
The derivative of $\mathcal{J}(\xk)$ with respect to the system state $\xk$ is given by
\begin{equation} \label{eqn:GF_Potential_Energy_Gradient}
  \nabla_\x \mathcal{J}(\xk) = \mathbf{B}^{-1}_k\, ( \xk - \xb_k ) - \mathbf{H}_k^T \, \mathbf{R}^{-1}_k\, \bigl( \y_k - \mathcal{H}_k(\x)\bigr)\, ,
\end{equation}
where $\mathbf{H}_k = \mathcal{H}_k'(\x) $ is the linearized observation operator (e.g. the Jacobian).  

The HMC sampling filter~\cite{attia2015hmcfilter}  proceeds in two steps, namely a forecast step and an analysis step. 
Given an analysis ensemble of states $\{ \xa_{k-1}{(e)}\}_{e=1,2,\ldots ,\nens}$ at time $t_{k-1}\,,$ an ensemble of forecasts at time $t_k$ is generated 
using the forward model $\mathcal{M}$:
\begin{equation} \label{eqn:forward_model_propagation}
\xb_k{(e)} = \mathcal{M}_{t_{k-1} \rightarrow t_k} \left( \xa_{k-1}{(e)} \right),\quad e=1,2,\ldots,\nens.
\end{equation}
In the analysis step, the posterior~\eqref{eqn:GF_Posterior} is sampled by running a HMC sampler with potential energy set to~\eqref{eqn:GF_Potential_Energy},
where $\mathbf{B}_k$ is approximated using the available ensemble of forecasts.

The formulation of the HMC filter presented in~\cite{attia2015hmcfilter}, and reviewed above, tends to be restrictive due to the assumption that the prior 
is always approximated by a Gaussian distribution. 
The prior distribution can be viewed as the result of propagating the posterior of the previous assimilation cycle using model dynamics.
In the case of nonlinear model dynamics, the prior distribution  is a nonlinear transformation of a non-Gaussian distribution which is generally expected to be non-Gaussian.
Tracking the prior distribution exactly however is not possible, and a relaxation assumption must take place.

% The main goal here is to present a more accurate representation of the prior that a Gaussian, given the forecast ensemble. 
% We propose fitting a Gaussian Mixture Model (GMM) to the prior ensemble, and sample the posterior formulated given this appproximation.
We propose conducting a more accurate density estimation of the prior, by fitting a GMM to the available prior ensemble, replacing the Gaussian prior with 
a Gaussian mixture prior.
% Specifically, we propose modifying the HMC sampling Filter by incorporating a clustering step in the forecast phase of the filter to relax the Gaussian prior assumption.
%
%

%~~~~~~~~~~~~~~~~~~~~~~~~~~~~~~~~~~~~~~~~~~~~~~~~~~~~~~~~~~~~~~~~~~~  
\section{Cluster Sampling Filters} \label{Sec:GMM_and_ClHMC}
%~~~~~~~~~~~~~~~~~~~~~~~~~~~~~~~~~~~~~~~~~~~~~~~~~~~~~~~~~~~~~~~~~~~  
    %

    %~~~~~~~~~~~~~~~~~~~~~~~~~~~~~~~~~~~~~~~~~~~~~~~~~~~~~~~~~~~~~~~~~~~  % Probably we don't need the subsection header here!
    \subsection{Mixture models} \label{Subsec:Mixture_Models}
    %~~~~~~~~~~~~~~~~~~~~~~~~~~~~~~~~~~~~~~~~~~~~~~~~~~~~~~~~~~~~~~~~~~~
%
The probability distribution $\PD(\x)$ is said to be a mixture of $\nc$ probability distributions $\{\mathcal{C}_i(\x)\}_{i=1,2,\ldots,\nc}$,
if $\PD(\x)$ takes the form:
\begin{equation} 
\label{eqn:mixture_distribution}
  \PD(\x) = \sum_{i=1}^{\nc}{\tau_i\, \mathcal{C}_i(\x) }\quad \textnormal{where} \quad 
  \tau_i > 0,~\forall i ~~ \textnormal{and}~~\sum_{i=1}^{\nc}{\tau_i} = 1\,.
\end{equation}
The weights $\tau_i$ are commonly referred to as the mixing weights, and $\mathcal{C}_i(\x)$ are the densities of the mixing components.
%
%______________________________________
\subsubsection{Gaussian mixture models (GMM)}
%______________________________________
%
A GMM is a special case of~\eqref{eqn:mixture_distribution} where the mixture components are Gaussian densities,  that is $\mathcal{C}_i(\x)=\mathcal{N}(\x;\Theta_i)$ with $\Theta_i=\{\mu_i,\,\Upsigmai\}$ being the parameters of the $i^{th}$ Gaussian component.

Fitting a GMM to a given data set is one of the most popular approaches for density estimation. Given a data set $\{\x(e)\}_{e=1,\,2,\,\ldots,\,\nens}\,,$ sampled from an unknown probability distribution $\PD(\x)$, one can estimate the density function $\PD(\x)$ by a GMM; the parameters of the GMM, i.e. the mixing weights $\tau_i$, the means $\mu_i$, and the covariances $\Upsigmai\,$ of the mixture components, can be inferred from the data.
     
The most popular approach to obtain the maximum likelihood estimate of the GMM parameters is the expectation-maximization (EM) algorithm~\cite{dempster1977maximum}.
% EM is a general procedure for parameter estimation, that is not limited to GMM parameters estimation, however we describe it only in our context.
EM is an iterative procedure that alternates between two steps, expectation (E) and maximization (M). 
At iteration $t+1$ the E-step computes the expectation of the complete log-likelihood based on the posterior probability of $\x$ belonging to the $i^{th}$ component, 
with the parameters $\Theta^{\{t\}}$ from the previous iteration.  
In particular, the following quantity $Q(\Uptheta| \Uptheta^{\{t\}})$ is evaluated:
\begin{equation}
  \begin{split} 
     \label{eqn:EM_QoI_ML}
Q(\Uptheta| \Uptheta^{\{t\}}) &= \sum_{e=1}^{\nens} \sum_{i=1}^{\nc} r_{e,i} \, \log{ \big(\tau_i \, \mathcal{N} \left( \x(e) ;\Theta_i \right) \big) }\,,  \\
r_{e,i}&= \frac{\tau_i^{\{t\}} \; \mathcal{N} \left(\x(e) ;\Theta_i^{\{t\}} \right) }{ \sum_{\ell=1}^{\nc} { \tau_\ell^{\{t\}} \; \mathcal{N} \left(\x(e) ;\Theta_\ell^{\{t\}} \right) } } \,,  \\
w_i    &= \sum_{e=1}^{\nens} r_{e,i} \,.  % w_i is the sum of probs per comp. Ahmed: Find an alternative symbol
  \end{split}
\end{equation}  
Here $\Uptheta=\{\tau_i,\Theta_i\}_{i=1 \dots \nc}$ is the parameter set of all the mixture components, and $r_{e,i}$ is the probability that the $e^{th}$ ensemble member lies 
under the $i^{th}$ mixture component.

In the M-step, the new parameters $\Uptheta^{\{t+1\}} =\arg\max_\Uptheta\, Q$ are obtained by maximizing the conditional probability $Q$ in \eqref{eqn:EM_QoI_ML} with respect to the parameters $\Uptheta$.  The updated parameters $\Uptheta^{\{t+1\}}$ are given by the analytical formulas: 
\begin{equation} \label{eqn:ME_Updated_Parameters}
  \begin{aligned}
    \tau_i^{\{t+1\}}     &= \frac{ \sum_{e=1}^{\nens}{r_{e,i}} }{ \nens } = \frac{w_i}{\nens} \,, \\
    \mu_i^{\{t+1\}}&= \sum_{e=1}^{\nens}{ \x(e) \, \frac{r_{e,i}}{w_i} } \,, \\
    \Upsigma_i^{\{t+1\}} &= \sum_{e=1}^{\nens} {  \left( \x(e)-\mu_i^{\{t+1\}} \right) \left( \x(e)-\mu_i^{\{t+1\}} \right)^T  \, \frac{r_{e,i}}{w_i} } \,.
  \end{aligned}
\end{equation}
To initialize the parameters for the EM iterations, the mixing weights 	are simply chosen to be equal $\tau_i=\nc^{-1}$,  the means $\mu_i$ can be randomly selected from the given ensemble,  and the covariance matrices of the components can be all set to covariance matrix of the full ensemble. 
% Regardless of the initialization, the convergence of the EM algorithm is ensured by the fact that it monotonically increases the log-likelihood at each iteration i.e. 
%
Regardless of the initialization, the convergence of the EM algorithm is ensured by the fact that it monotonically increases the observed data log-likelihood at each iteration~\cite{dempster1977maximum}, that is:
\begin{equation*} \label{eqn:EM_indirect_convergence_criteria}
    \sum_{e=1}^{\nens}{\log{ \left( \sum_{i=1}^{\nc}{ \tau_i^{\{t+1\}} \, \mathcal{N}\left(\x(e);\Theta_i^{\{t+1\}}\right)} \right)} } 
	\geq \sum_{e=1}^{\nens}{\log{ \left( \sum_{i=1}^{\nc}{ \tau_i^{\{t\}} \, \mathcal{N}\left(\x(e);\Theta_i^{\{t\}}\right)} \right)} }\,.
\end{equation*}
EM algorithm achieves the improvement of the data log-likelihood indirectly by improving the quantity $Q(\Uptheta| \Uptheta^{\{t\}})$ over consecutive iterations, i.e. $Q(\Uptheta| \Uptheta^{\{t+1\}}) \geq Q(\Uptheta| \Uptheta^{\{t\}})$.

%______________________________________
\subsubsection{Model selection}
%______________________________________
%
Before EM iterations start, the number of mixture components $\nc$ must be detected. To choose the number of components in the prior mixture model selection is employed. Model selection is a process of selecting a model in the set of a candidate models that gives the best trade-off between model fit and complexity. 
Here, the best number of components $\nc$ can be selected with common model selection methodologies such as Akaike Information Criterion (AIC) and Bayesian Information Criterion (BIC): 
  \begin{equation} \label{eqn:aic_bic}
    \begin{aligned}
      % AIC (\mathcal{M}) &= -2\, \sum_{e=1}^{\nens}  \log C \left( \x(e) ;\widehat{\Theta} \right) +2 N_{\Theta}\,,  \\
      % AIC &= -2\, \sum_{e=1}^{\nens}  \log\, \mathcal{C} \left( \x(e) ;\widehat{\Theta} \right) + { 2\, (3 \nc - 1) } \,,  \\
      AIC &= -2\, \sum_{e=1}^{\nens}{\log{ \left( \sum_{i=1}^{\nc}{ \widehat{\tau}_i \, \mathcal{N}\left(\x(e);\widehat{\Theta}_i\right)} \right)} }  + { 2\, (3 \nc - 1) } \,,  \\  % <-- Ahmed
      % BIC(\mathcal{M})  &= -2\, \sum_{e=1}^{\nens}  \log C \left( \x(e) ;\widehat{\Theta}\right) + N_{\Theta} \log (\nens) \,,
      % BIC &= -2\, \sum_{e=1}^{\nens}  \log \,\mathcal{C} \left( \x(e) ;\widehat{\Theta}\right) + {  \log (\nens) \,(3 \nc - 1)} \,,
      BIC &= -2\, \sum_{e=1}^{\nens}{\log{ \left( \sum_{i=1}^{\nc}{ \widehat{\tau}_i \, \mathcal{N}\left(\x(e);\widehat{\Theta}_i\right)} \right)} } + {  \log (\nens) \,(3 \nc - 1)} \,,  % <-- Ahmed
    \end{aligned}
\end{equation}
%
% where $\widehat{\Theta}$ is the set of optimal model parameters, $ C \left( \x(e) ;\widehat{\Theta} \right) $ is the maximized value of the likelihood function of the candidate GMM model with $\nc$ components.
where $\{\widehat{\tau}_i,\,\widehat{\Theta}_i\}_{i=1 \dots \nc}$ is the set of optimal parameters for the candidate GMM model with $\nc$ components.
%
% where $\widehat{\Theta}$ is the set of optimal model parameters, $ C \left( \x(e) ;\widehat{\Theta} \right) $ is the maximized value of the likelihood function of 
% the candidate GMM model, and $N_{\Theta}$  is the number of estimated parameters in the candidate GMM model with $\nc$ components. 
% We can leave N{\Theta} as in the old (commented) version of the equation, and explain it as foloows:
% The number of parameters in the GMM:
% Consider a candidate GMM with \nc compoenents; the number of model parameters $N_{\Theta}$ is:
% Mixing weights: \nc -1  % since sum of weights is equal to on, we have only \nc-1 degrees of freedom here.
% Mixture means: \nc
% Mixtuer Covariances: \nc
% N_{\Theta} = 3 \nc - 1

The best number of components $\nc$ minimizes the AIC or BIC criterion~\cite{hu2004investigation,schwarz1978estimating}. 
% The main difference between the two criteria is that BIC imposes greater penalty for the number of parameters. 
The main difference between the two criteria, as explained by the second terms in Equation~\eqref{eqn:aic_bic}, is that BIC imposes greater penalty on the number of parameters ($3 \nc -1$) of the candidate GMM model.
For small or moderate numbers of samples  BIC often chooses models that are too simple because of its heavy penalty on complexity. 
% BIC is more consistent as a selection criterion, it has a property that given a  family of models including the true model, the probability that BIC will 
% select the correct one approaches one as the sample size becomes large while AIC tends to choose more complex models as the number of sample size increases.
%
%

%~~~~~~~~~~~~~~~~~~~~~~~~~~~~~~~~~~~~~~~~~~~~~~~~~~~~~~~~~~~~~~~~~~~
\subsection{Cluster HMC sampling filter (\ClHMC)} 
\label{subsec:cluster_HMC_filter}
%~~~~~~~~~~~~~~~~~~~~~~~~~~~~~~~~~~~~~~~~~~~~~~~~~~~~~~~~~~~~~~~~~~~
%
The prior distribution is approximated by a GMM fitted to the forecast ensemble, e.g., using an EM clustering step. The prior PDF reads:
\begin{equation} \label{eqn:GMM_Filter_Prior}
\Pb(\xk) = \sum_{i=1}^{\nc}{\tauki\, \mathcal{N}(\x;\muki,\, \Upsigmaki) }\,  
 = \sum_{i=1}^{\nc}{\tauki\,  \frac{(2\pi)^{-\frac{\nvar}{2}}}{\sqrt{|\Upsigmaki|}}\, 
  \exp{\left( - \frac{1}{2}  \lVert \x-\muki \rVert^2_ {\mathbf{\Upsigmaki}^{-1}}  \right) } }\,,
\end{equation}
where the weights $\tauki$ quantify the probability that an ensemble member $\xk(e)$ belongs to the $i^{th}$ component, and $(\muki,\, \Upsigmaki)$
are the mean and the covariance matrix associated with the $i^{th}$ component of the mixture model at time instance $t_k$.
%
% The number of mixture components $\nc$ can be inferred from the ensemble using an information criterion such as AIC, or BIC as described in Subsection~\ref{Subsec:Model_Selection}.
% The mean and the covariance matrix of each component can be estimated using the members belonging to this cluster in the ensemble.

Assuming Gaussian observation errors, the posterior can be formulated using equations \eqref{eqn:Bayes_Filtering_Rule}, \eqref{eqn:GF_Likelihood}, and \eqref{eqn:GMM_Filter_Prior} as follows: 
\begin{equation} 
\label{eqn:ClHMC_Filter_Posterior}
\begin{split}
f(\xk) &= \Pa(\xk)   \\ %= \PD(\xk|\yk) 
 &= \frac{(2\pi)^{-\frac{\Nobs}{2}}}{\sqrt{|\mathbf{R}_k|}}\, 
\exp{\left( -\frac{1}{2}  \lVert \mathcal{H}_k(\xk) -\yk \rVert^2 _{\mathbf{R}_k^{-1} } \right) }\,
\sum_{i=1}^{\nc}{\tauki\,  \frac{(2\pi)^{-\frac{\nvar}{2}}}{\sqrt{|\Upsigmaki|}}\, 
\exp{\left( - \frac{1}{2}  \lVert \xk-\muki \rVert^2_ {\mathbf{\Upsigmaki}^{-1}}  \right) } }  \\
 &\propto  \phi(\xk) = \sum_{i=1}^{\nc}{\frac{\tauki}{\sqrt{|\Upsigmaki|}}\, 
\exp{\left( -\frac{1}{2} \lVert \xk-\muki \rVert^2_{\mathbf{\Upsigmaki}^{-1}}
    -\frac{1}{2} \lVert \mathcal{H}_k(\xk) - \yk \rVert^2_{\mathbf{R}_k^{-1}} \right) }}\,.
\end{split}
\end{equation}
In general the posterior PDF \eqref{eqn:ClHMC_Filter_Posterior} will not correspond to a Gaussian mixture due to the nonlinearity of the observation operator. This makes analytical solutions not possible. Here we seek to sample directly from the posterior PDF \eqref{eqn:ClHMC_Filter_Posterior}. 
%Our hope is that the sampling filter will be able to capture the structure of the posterior via pure sampling.

The HMC sampling requires setting the potential energy term in the Hamiltonian~\eqref{eqn:Hamiltonian_function} to the negative-log of the 
posterior distribution \eqref{eqn:ClHMC_Filter_Posterior}. The potential energy term $\mathcal{J}(\xk)$ is:
\begin{subequations} \label{eqn:ClHMC_Filter_Potential_Energy_pre}
\begin{equation}
\begin{aligned}
\mathcal{J}(\xk) &= -\log{ \left(  \sum_{i=1}^{\nc}{\frac{\tauki}{\sqrt{|\Upsigmaki|}}\, 
\exp{\left( -\frac{1}{2} \lVert \xk-\muki \rVert^2_{\mathbf{\Upsigmaki}^{-1} }
    -\frac{1}{2} ( \lVert \mathcal{H}_k(\xk) - \yk\rVert^2_{ \mathbf{R}_k^{-1}} \right) }}  \right)  }   \\
 &= \frac{1}{2} \lVert \mathcal{H}_k(\xk)-\yk \rVert^2_ {\mathbf{R}_k^{-1}} -\log{ \left(  \sum_{i=1}^{\nc}{\frac{\tauki}{\sqrt{|\Upsigmaki|}}\, 
\exp{\left( -\frac{1}{2} \lVert \xk-\muki \rVert^2_{\mathbf{\Upsigmaki}^{-1}} \right) }}  \right)  }  \\
 &= \frac{1}{2} \lVert \mathcal{H}_k(\xk)-\yk \rVert^2_{\mathbf{R}_k^{-1}} -\log{ \left(  \sum_{i=1}^{\nc}{\frac{\tauki}{\sqrt{|\Upsigmaki|}}\, 
\exp{\big( - \mathcal{J}_{k,i}(\xk)  \big) }}  \right)  } \,,
\end{aligned}
\end{equation}
where
\begin{equation}
  \mathcal{J}_{k,i}(\xk) =  \frac{1}{2} \lVert \xk-\muki \rVert^2_{\mathbf{\Upsigmaki}^{-1}} \,. 
\end{equation}
\end{subequations}

Equation~\eqref{eqn:ClHMC_Filter_Potential_Energy_pre} is expected to suffer from numerical difficulties due to evaluating the logarithm of a sum of very small values. To address the accumulation of roundoff errors, and without loss of generality, we assume from now on that the terms in Equation~\eqref{eqn:ClHMC_Filter_Potential_Energy_pre} under the sum are sorted in decreasing order, i.e. 
$(\tauki/\sqrt{|\Upsigmaki|})\, \exp{\left(-\mathcal{J}_{k,i}(\xk) \right) } \,
>\, (\tau_{k,i+1}/\sqrt{|\mathbf{\Upsigma}_{k,i+1}|})\,\exp{\left(-\mathcal{J}_{k,i+1}(\xk) \right) }$, $\forall\, i=1,\ldots,\nc-1$. 

The potential energy function~\eqref{eqn:ClHMC_Filter_Potential_Energy_pre} is rewritten as:
\begin{subequations} 
\label{eqn:ClHMC_Filter_Potential_Energy}
\begin{align}
\mathcal{J}(\xk) 
  &= \frac{1}{2} \lVert \mathcal{H}_k(\xk) - \yk \rVert^2_{\mathbf{R}_k^{-1}}   \\
  & \hspace{1cm} - \left[   
  \log{ \left( \frac{\tau_{k,1}\, 
\exp{\left( - \mathcal{J}_{k,1}(\xk)  \right) }}{\sqrt{|\mathbf{\Upsigma}_{k,1}|}} \right) 
}
+ \log{ \left( 1 + \sum_{i=2}^{\nc}{ \frac{  \frac{\tauki}{\sqrt{|\Upsigmaki|}}\, 
\exp{\left( - \mathcal{J}_{k,i}(\xk)  \right) }  }{\frac{\tau_{k,1}}{\sqrt{|\mathbf{\Upsigma}_{k,1}|}}\, 
\exp{\left( - \mathcal{J}_{k,1}(\xk)  \right) } }  }  \right)  
} 
  \right]  \nonumber   \\
  &= \frac{1}{2} \lVert \mathcal{H}_k(\xk) - \yk \rVert^2_{\mathbf{R}_k^{-1}}  + \mathcal{J}_{k,1}(\xk) \\
  & \hspace{1cm} - \log{ \left( \frac{\tau_{k,1}}{\sqrt{|\mathbf{\Upsigma}_{k,1}|}} \right) }    \nonumber 
- \log{ \left( 1+  \sum_{i=2}^{\nc}{  \frac{\tauki \, \sqrt{|{\Upsigma}_{k,1}|} }{\tau_{k,1} \, \sqrt{|\Upsigmaki|}}\, 
\exp{ \big( \mathcal{J}_{k,1}(\xk)  - \mathcal{J}_{k,i}(\xk)  \big) } } \right)  
    }. \nonumber     
\end{align}
\end{subequations}
The gradient of the potential energy~\eqref{eqn:ClHMC_Filter_Potential_Energy} is:
\begin{subequations} 
\label{eqn:ClHMC_Filter_Potential_Energy_Gradient}
\begin{align}
    \nabla_{\xk}\mathcal{J}(\xk) 
&= \mathbf{H}_k^T \mathbf{R}_k^{-1} ( \mathcal{H}_k(\xk) - \yk ) + \nabla_{\xk} \mathcal{J}_{k,1}(\xk)   \\
& \quad - \frac{1}{ \left( 1 + \sum_{i=2}^{\nc}{  \frac{\tau_{k,i} \, \sqrt{|\mathbf{\Upsigma}_{k,1}|} }
{\tau_{k,1} \, \sqrt{|\mathbf{\Upsigma}_{k,i}|} } \, 
   \exp\big( \mathcal{J}_{k,1}(\xk) - \mathcal{J}_{k,i}(\xk)  \big) }  
    \right) 
  }  \nonumber \\
& \hspace{1cm} \cdot \sum_{i=2}^{\nc} \Big\{ \frac{\tauki \, \sqrt{|\Upsigma_{k,1}|} }{\tau_{k,1} \, \sqrt{|\Upsigmaki|}}\, 
   \exp\big( \mathcal{J}_{k,1}(\xk) - \mathcal{J}_{k,i}(\xk)  \big) \cdot \nonumber \\
& \hspace{2cm} \cdot    \Bigl[ \nabla_{\xk}\mathcal{J}_{k,1}(\xk) - \nabla_{\xk}\mathcal{J}_{k,i}(\xk) \Bigr] \Big\}
  \,, \nonumber     
\end{align}
%
% where $\mathbf{H}_k$ is the linearized observation operator at time $t_k\,,$ and $\nabla_{\xk} \mathcal{J}_{k,i}(\xk)$ is the derivative of the $i^{th}$ component of the mixture and is given by:
where $\nabla_{\xk} \mathcal{J}_{k,i}(\xk)$ is given by:
\begin{equation} \label{eqn:ClHMC_Filter_GMM_Component_Dervative}
    \nabla_{\xk} \mathcal{J}_{k,i}(\xk) = \Upsigmaki^{-1}(\xk-\muki)\,\, \quad \forall\, i=1,2,\ldots,\nc.
\end{equation}     
\end{subequations}

In the case where the mixture contains a single component (one Gaussian distribution) the potential  energy function~\eqref{eqn:ClHMC_Filter_Potential_Energy} and it's gradient~\eqref{eqn:ClHMC_Filter_Potential_Energy_Gradient} reduce to the following, respectively:
\begin{equation}
\begin{split}
\mathcal{J}(\xk) &= \frac{1}{2} \lVert \xk - \xb_k \rVert^2_{\mathbf{B}_k^{-1}} + 
    \frac{1}{2} \lVert \mathcal{H}_k(\xk) - \yk \rVert^2_{\mathbf{R}_k^{-1}}\,, \\
\nabla_{\xk}\mathcal{J}(\xk) &=  \mathbf{B}_k^{-1} \left( \xk - \xb_k \right) + \mathbf{H}_k^T \mathbf{R}_k^{-1} \left( \mathcal{H}_k(\xk) - \yk \right) \,.
\end{split}
\end{equation}
This shows that the \ClHMC sampling filter proposed herein reduces to the original HMC filter the EM algorithm detects a single component  during the prior density approximation phase.

%______________________________________
\paragraph{\bf The \ClHMC sampling algorithm}
%______________________________________
%
%Having formulated the potential energy function~\eqref{eqn:ClHMC_Filter_Potential_Energy} and it's gradient~\eqref{eqn:ClHMC_Filter_Potential_Energy_Gradient}, 
%we can now detail the steps of the \ClHMC sampling filter.  

As in the HMC sampling filter, information about the analysis probability density at the previous time $t_{k-1}$ is captured by the analysis ensemble of 
states $\{ \xa_{k-1}{(e)}\}_{e=1,\ldots ,\nens}$. The forecast step consists of two stages. First, the model \eqref{eqn:forward_model_propagation} is used to integrate each analysis ensemble member forward to time $t_{k}$ where observations are available. Next, a clustering scheme (e.g., EM) is used to generate the parameters of the GMM. The analysis step constructs a Markov chain starting from an initial state $\xk^{0}$, and proceeds by sampling the posterior PDF~\eqref{eqn:ClHMC_Filter_Posterior} at stationarity.  Here the superscript over $\xk$ refers to the iteration number in the Markov chain.

The steps of the \ClHMC sampling filter are detailed in Algorithm \ref{Alg:ClHMC_sampling_filter}. As discussed in~\cite{attia2015hmcfilter},  Algorithm~\ref{Alg:ClHMC_sampling_filter} can be used either as a non-Gaussian filter, or as a replenishment tool  for parallel implementations of the traditional filters such as EnKF.
\begin{algorithm}[htbp!]
\begin{algorithmic}[1]
  \STATE \textbf{Forecast step:} given an analysis ensemble $\{ \xa_{k-1}{(e)}\}_{e=1,2,\ldots ,\nens}$ at time $t_{k-1}$;
    \begin{enumerate}
\item[i-  ] generate the forecast ensemble using the model $\mathcal{M}$:
  \begin{equation}
  \xb_k{(e)} = \mathcal{M}_{t_{k-1} \rightarrow t_k} \left( \xa_{k-1}{(e)} \right),\quad e=1,2,\ldots,\nens. \nonumber
  \end{equation}
\item[ii- ] Use AIC/BIC criteria to detect the number of mixture components $\nc$ in the GMM, then
  use EM to estimate the GMM parameters $\{(\tauki;\,\muki,\, \Upsigmaki)\}_{i=1,2,\ldots,\nc}$.
    \end{enumerate}
  \STATE \textbf{Analysis step:} given the observation $\yk$ at time point $t_k\,,$ follow the steps $i$ to $v$:
    \begin{enumerate}
\item[i-  ] Initialize the Markov Chain ($\xk^{0}$) to be to the best estimate available, 
  e.g. to the mean of the joint forecast ensemble, or the mixture component mean with maximum likelihood.
  %An initial state closer to the posterior leads to faster convergence of the chain.
%
\item[ii- ] Choose a positive definite mass matrix $\mathbf{M}$. 
  A recommended choice is to set $\mathbf{M}$ to be a diagonal matrix whose diagonal is equal to the diagonal of the posterior precision matrix.
  The precisions calculated from the prior ensemble can be used as a proxy.
\item[iii-] %Apply Algorithm \ref{alg:HMC_sampling} with initial state $ \xk^{0}$ and generate $\nens$ ensemble members. 
  Set the potential energy function to~\eqref{eqn:ClHMC_Filter_Potential_Energy}, and it's derivative 
  to~\eqref{eqn:ClHMC_Filter_Potential_Energy_Gradient}.
\item[iv-]  Initialize the chain with a state $\xk^{0}$ and generate $\nens$ ensemble members from the posterior 
  distribution~\eqref{eqn:ClHMC_Filter_Posterior} as follows:
  \begin{enumerate} [1)]
    \item  Draw a random vector $\p^{r} \sim \mathcal{N}(0,\mathbf{M})$.
    \item Use a symplectic numerical integrator (e.g. Verlet, 2-Stage, or 3-Stage~\cite{sanz2014Markov,attia2015hmcfilter}) to advance 
  the current state $(\p^{r},\, \xk^{r} )$ by a pseudo-time increment $T$ to obtain a \textit{proposal} state $( \p^{*},\, \xk^{*} )$: 
  \begin{equation}
( \p^*,\, \xk^{*} ) = \Phi_T\bigl(( \p^r,\, \xk^{r} )\bigr).
  \end{equation}
    \item Evaluate the energy loss :
  $
\Delta H =  H( \p^*,\, \xk^{*} ) - H( \p^r,\, \xk^{r} ).
  $
    \item Calculate the acceptance probability:
  $
a^{(r)} = 1 \wedge e^{-\Delta H}.
  $
    \item Discard both $\p^{*},\, \p^{r}$.
    \item \textbf{(Acceptance/Rejection)} Draw a uniform random variable $u^{(r)}\sim \mathcal{U}(0,1)$:
  \begin{enumerate}
  \item[i-  ] If $a^{(r)} > u^{(r)}$ accept the proposal as the next sample: $\xk^{r+1} := \xk^{*}$;
  \item[ii- ] If $a^{(r)} \leq u^{(r)}$ reject the proposal and continue with the current state: $\xk^{r+1} := \xk^{r}$.
  \end{enumerate}
    \item Repeat steps $1$ to $6$ until $\nens$ distinct samples are drawn.
  \end{enumerate}
\item[v- ]  Use the generated samples $\{ \xa_k(e)\}_{e=1,2,\ldots ,\nens}$ as an analysis ensemble. 
  The analysis ensemble can be used to infer the posterior moments, e.g. posterior mean and posterior covariance matrix.
    \end{enumerate}
  \STATE Increase time $k := k+1$ and repeat steps 1 and 2.
\end{algorithmic}
\caption{Cluster HMC sampling filter (\ClHMC)}
\label{Alg:ClHMC_sampling_filter}
\end{algorithm}
%
%

%~~~~~~~~~~~~~~~~~~~~~~~~~~~~~~~~~~~~~~~~~~~~~~~~~~~~~~~~~~~~~~~~~~~
\subsection{Computational considerations} 
\label{Subsec:Computational_considerations}
%~~~~~~~~~~~~~~~~~~~~~~~~~~~~~~~~~~~~~~~~~~~~~~~~~~~~~~~~~~~~~~~~~~~
%
To initialize the Markov chain one seeks a state that is likely with respect to the analysis distribution. Therefore one can start with the background ensemble mean, or with the mean of the component that has the highest weight. Alternatively, one can apply a traditional EnKF step and use the mean analysis to initialize the chain.

The joint ensemble mean and covariance matrix can be evaluated using the forecast ensemble, or using the GMM parameters.  Given the GMM parameters $(\tauki;\,\muki,\, \Upsigmaki)$, the joint background mean and covariance matrix are, respectively:
\begin{subequations} \label{eqn:Joint_mean_and_Covar}
\begin{eqnarray} 
  {\bf \overline{x}}^{\rm b}_k &=& \sum_{i=1}^{\nc}{\tauki \, \muki} \,, \label{eqn:Joint_mean} \\
  \mathbf{B}^{\rm ens}_k&=& \sum_{i=1}^{\nc}{\tauki \, \Upsigmaki} + \sum_{i=1}^{\nc}{\tauki \,
    % \lVert \muki - {\bf \overline{x}}^{\rm b}_k \rVert ^2_{\mathbf{I}} \,, \label{eqn:Joint_Covar}
    (\muki - {\bf \overline{x}}^{\rm b}_k) (\muki - {\bf \overline{x}}^{\rm b}_k)^T \,. 
    }
\end{eqnarray}
\end{subequations}

Both the potential energy~\eqref{eqn:ClHMC_Filter_Potential_Energy} and it's gradient~\eqref{eqn:ClHMC_Filter_Potential_Energy_Gradient} require evaluating the determinants of the covariance matrices associated with the mixture components.  This is a computationally expensive process that is best avoided for large-scale problems. A simple remedy is to force the covariance matrices $\Upsigma_{k,i},\, \forall i = 1,2,\ldots,\nc$ to be diagonal while constructing the GMM.

When the Algorithm~\ref{Alg:ClHMC_sampling_filter} is applied sequentially at some steps a single mixture component could be detected in the 
prior ensemble. In this case forcing a diagonal covariance structure does not help; in this case the ensemble covariance is calculated and the standard HMC sampler step is applied.

\subsection{A multi-chain version of the \ClHMC ~filter (MC-\ClHMC) } 
\label{subsec:MC_ClHMC}
%~~~~~~~~~~~~~~~~~~~~~~~~~~~~~~~~~~~~~~~~~~~~~~~~~~~~~~~~~~~~~~~~~~~
%
%
Given the special geometry of the posterior mixture distribution one can construct separate Markov chains for different components of the posterior.  These chains can run in parallel to  independently sample different regions of the analysis distribution. By running a Markov chain starting at each component of the mixture distribution we ensure that the proposed algorithm navigates all modes of the posterior, and covers all regions of high probability. 

The parameters of the jumping distribution for each of the chains can be tuned locally based on the statistics of the ensemble points  belonging to the corresponding component in the mixture. This approach is potentially very efficient, not only because it reduces the total running time of the sampler, but also because it favors an increase acceptance rate. The local ensemble size (sample size per chain) can be specified based on the prior weight of the corresponding component  multiplied by the likelihood of the mean of that component. Every chain is initialized to the mean of the corresponding component in the prior mixture. The diagonal of the mass matrix can be set globally for all components, for example using the diagonal of the precision matrix of the forecast ensemble, or can be chosen locally based on the second-order moments estimated from the prior ensemble under the corresponding component in the prior mixture. This local choice of the mass matrix does not change the marginal density of the target variable.

%The sampler starts by running an EM step to build a GMM approximation of the prior distribution using the given forecast ensemble.
% Assuming EM is run sequentially, once the GMM is constructed on the root processor, the GMM information is broadcasted to all the working nodes. 
% Of course, if the number of processors is exactly the same as the number of components, each node is assigned one chain. 

% In the next section, we test the \ClHMC, and the MC-\ClHMC filters against both the traditional HMC filter, and EnKF using QG model.
%The two filters \ClHMC, and MC-\ClHMC are tested against the traditional HMC sampling filter and the EnKF in the next Section~\ref{Sec:Numerical_Results} 
%using a quasi-geostrophic model.
%
%
  %
  %____________________________________________________________________________
  \section{Numerical Results} 
  \label{Sec:Numerical_Results}
  %____________________________________________________________________________
    %
    %
We first apply the proposed algorithms, \ClHMC and MC-\ClHMC to sample a simple one-dimensional mixture distribution. 
The proposed methodologies are then tested using a quasi-geostrophic (QG) model and compared against the  original HMC sampling filter and against EnKF.  We mainly use a nonlinear 1.5-layer reduced-gravity QG model with double-gyre wind forcing and bi-harmonic friction~\cite{sakov2008deterministic}.

%~~~~~~~~~~~~~~~~~~~~~~~~~~~~~~~~~~~~~~~~~~~~~~~~~~~~~~~~~~~~~~~~~~~
\subsection{One-dimensional test problem} 
\label{subsec:OneD_Examplel}
\begin{figure}[!htbp]
\centering
\includegraphics[width=0.47\linewidth]{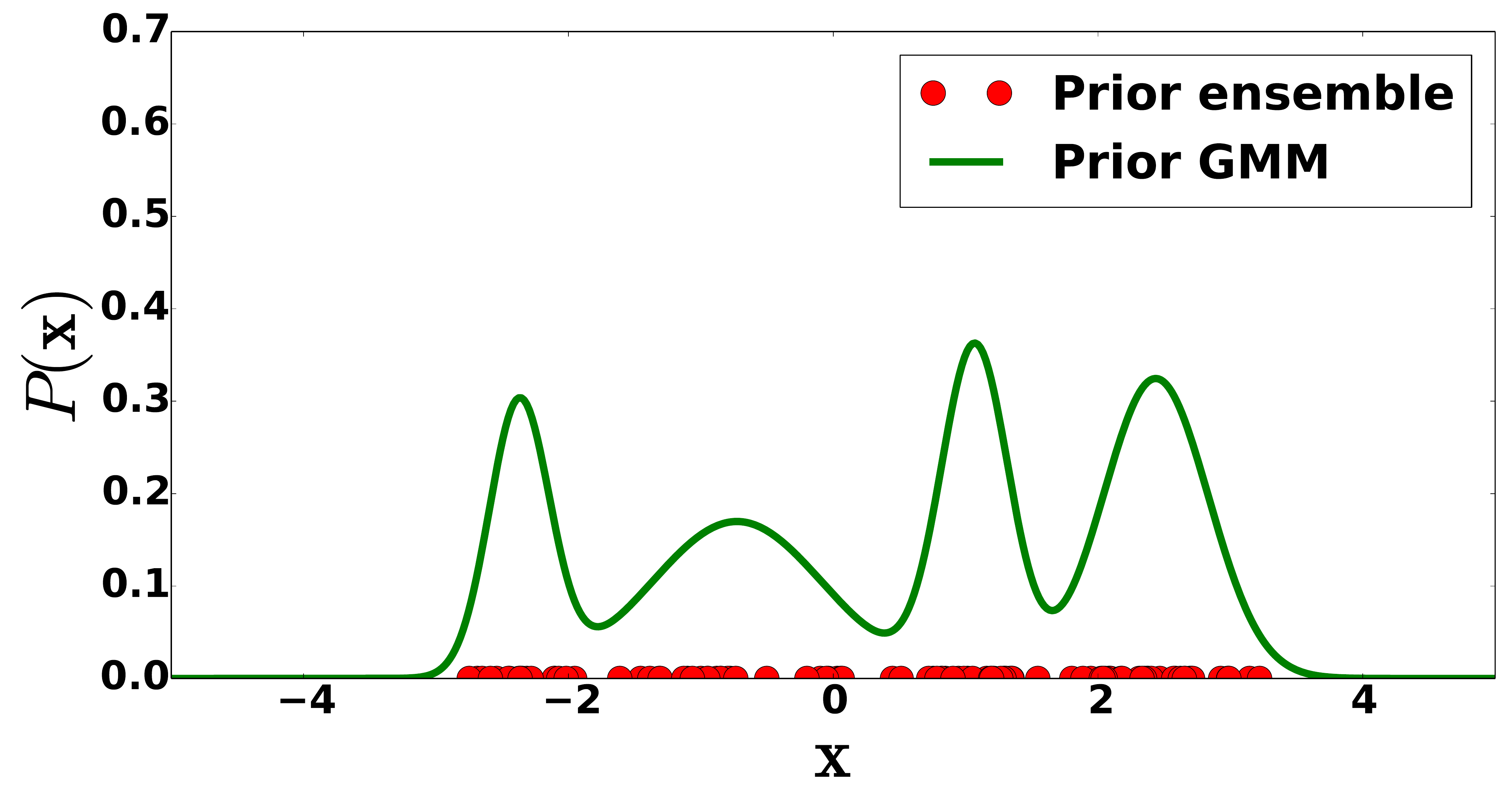}
\caption{The one-dimensional example. A random sample of size $\nens=100$ generated from a GMM with parameters 
	 given by~\eqref{eqn:OneD_Example_true_gmm_parameters}, and a GMM constructed by EM algorithm with AIC model selection criterion. 
	 }
\label{fig:OneD_example_prior_ensemble}
\end{figure}

\begin{figure}[!htbp]
\centering
\subfigure[\ClHMC]{
\includegraphics[width=0.47\linewidth]{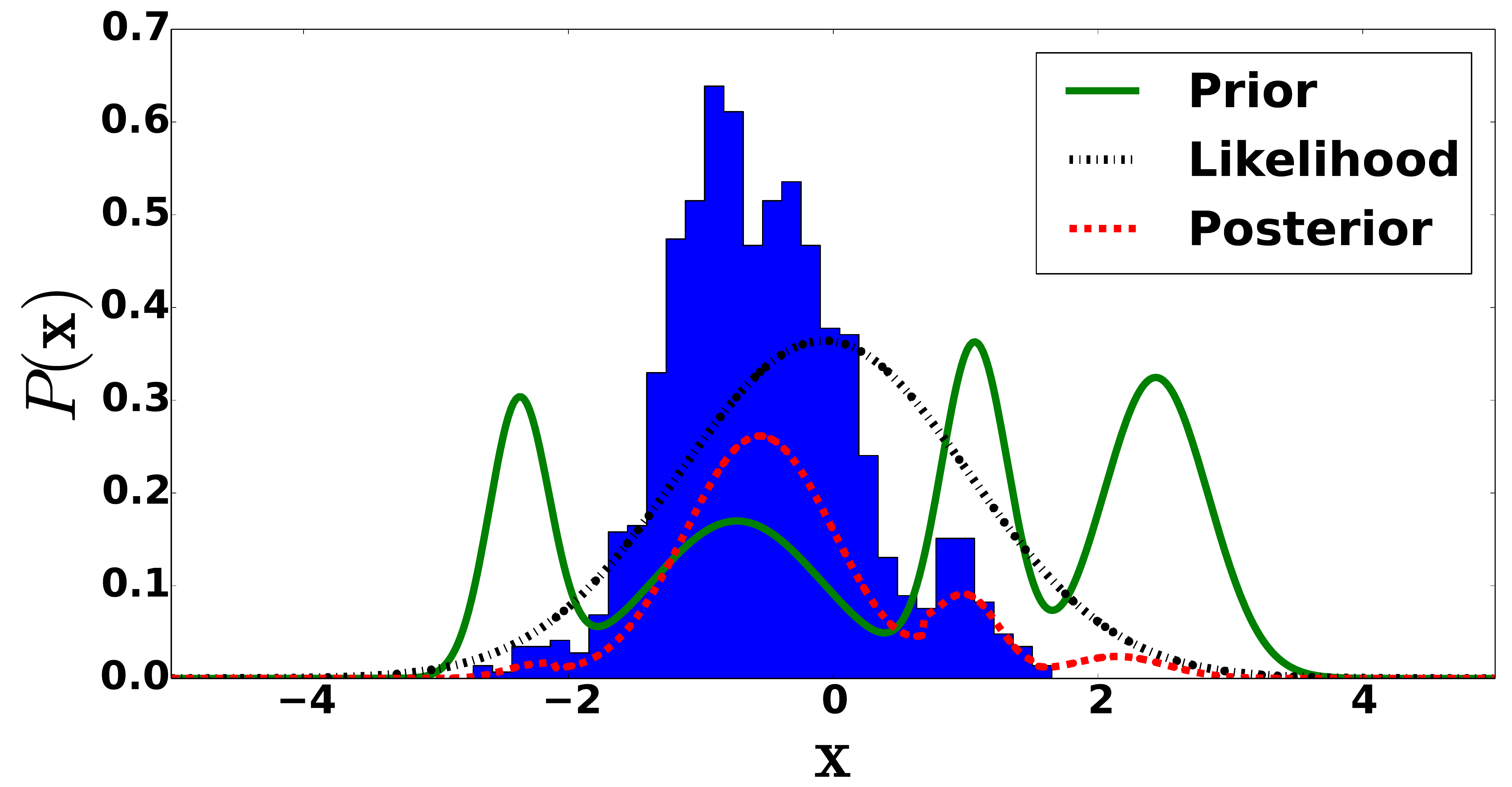}
\label{fig:OneD_sampling_example_serial}
}
\hfill
\subfigure[MC-\ClHMC]{
\includegraphics[width=0.47\linewidth]{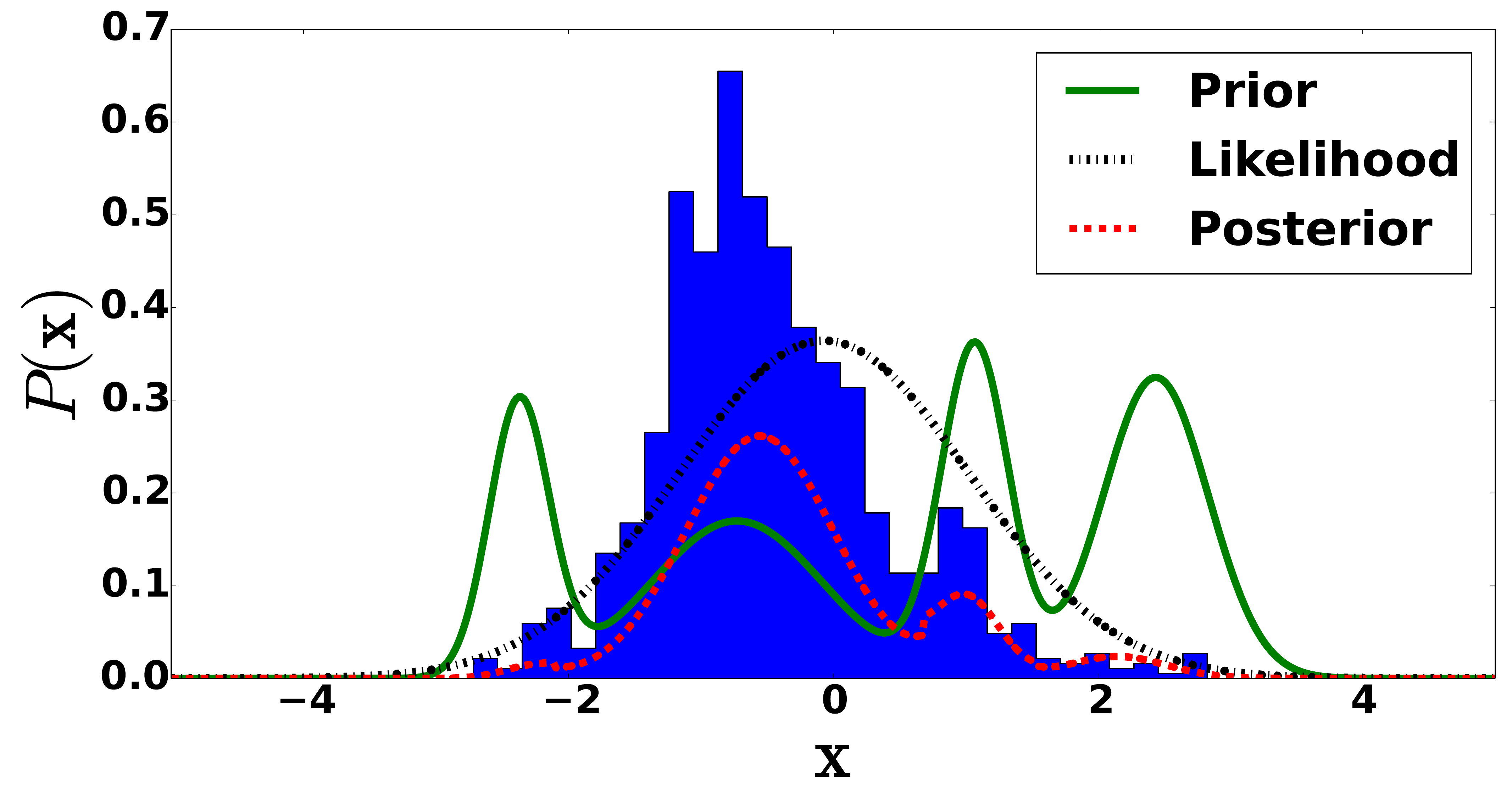}
\label{fig:OneD_sampling_example_parallel}
}
\caption{The one-dimensional example. A GMM, a Gaussian likelihood, and the resulting posterior, along with histograms of $1000$ sample points generated by the \ClHMC, and the MC-\ClHMC sampling algorithms. The sampling schemes are indicated under each panel. The symplectic integrator used is Verlet with pseudo-time stepping parameters $T=mh$ with $m=20$, and $h=0.045$. The number of burn-in steps is zero, and the number of mixing steps is $15$.
 }
\label{fig:OneD_sampling_example}
\end{figure}

%~~~~~~~~~~~~~~~~~~~~~~~~~~~~~~~~~~~~~~~~~~~~~~~~~~~~~~~~~~~~~~~~~~~
%
% Before presenting the results of the proposed methodologies with the QG model, we start with testing the capability of the proposed algorithms to sample a simple One-dimensional mixture distribution.
We start with a prior ensemble generated from a GMM with $\nc=5$ and the following mixture parameters: 
\begin{equation} \label{eqn:OneD_Example_true_gmm_parameters}
\begin{split}
  \{(\tau_i;\,\mu_i, \sigma_i^2)\}_{i=1,\dots,5}=\{(0.2;\, -2.4, 0.05),\, (0.1;\, -1.0, 0.07),\, (0.1;\, 0, 0.02),\,\\
(0.3;\, 1.0, 0.06),\, (0.3;\, 2.4, 0.1) \, \} \,.
\end{split}
\end{equation}
The EM algorithm is used to construct a GMM approximation of the true probability distribution from which the given prior ensemble is drawn.
The model selection criterion used here is AIC.
% AIC is used to find the number of compoenents given the prior ensemble.
%
The generated GMM approximation of the prior has  $\nc=4$ and the following parameters:
\begin{equation} \label{eqn:OneD_Example_gmm_parameters}
\begin{split}
  \{(\tau_i;\,\mu_i, \sigma_i^2)\}_{i=1,\dots,4}=\{(0.169;\, -2.370, 0.052),\, (0.278;\, -0.727, 0.423),\\ 
     (0.229;\, 1.070, 0.065),\, (0.324;\, 2.436, 0.159) \,\} \,.
\end{split}
\end{equation}
The prior ensemble and the GMM approximation of the true prior are shown in Figure~\ref{fig:OneD_example_prior_ensemble}.
Assuming the observation likelihood function is given by:
\begin{equation}
\PD(\y|\x) =\frac{1}{\sqrt{1.2}\, \sqrt{ 2\pi }}\, 
\,\exp{\Bigl( -\frac{1}{2}\, \frac{\left( \x -\y  \right)^2}{ 1.2 } \Bigr) }\,,
\end{equation}
with an observation $\y=-0.06858$, the posterior and the histograms of $1000$ sample points generated by \ClHMC, and MC-\ClHMC algorithms, are shown in Figure~\ref{fig:OneD_sampling_example}. In this example, the symplectic integrator used is Verlet with pseudo-time stepping parameters $T=mh$ with $m=20$, and $h=0.05$. Since the chains are initialized to  the means of the prior mixture components, the burn-in stage is waived, i.e.. the number of burn-in steps is set to zero. To reduce the correlation between the ensemble members of one chain we discard $15$ states (mixing steps) between each two consecutive sampled points. In the MC-\ClHMC filter, the ensemble size per component (per chain) is set to $\nens \times \ell_i \times \tau_i $, where $ \ell_i$ is the likelihood of the mean of the $i^{\rm th}$ component in the prior mixture.
The results reported in Figures~\ref{fig:OneD_sampling_example_serial} and~\ref{fig:OneD_sampling_example_parallel} show that both \ClHMC and MC-\ClHMC algorithms are 
capable  of generating ensembles with mass distribution accurately representing the underlying target posterior.
\ClHMC however fails to sample one of the probability modes, while MC-\ClHMC generates samples from the vicinities of all posterior probability modes.

An implementation of the \ClHMC and MC-\ClHMC sampling algorithms is available from~\cite{ClHMC_1D_test_cases}.  
%

%~~~~~~~~~~~~~~~~~~~~~~~~~~~~~~~~~~~~~~~~~~~~~~~~~~~~~~~~~~~~~~~~~~~
\subsection{Quasi-geostrophic model} 
\label{subsec:QG_Model}
%~~~~~~~~~~~~~~~~~~~~~~~~~~~~~~~~~~~~~~~~~~~~~~~~~~~~~~~~~~~~~~~~~~~
%
We employ the QG-1.5 model described by Sakov and Oke~\cite{sakov2008deterministic}. This model is a numerical approximation of the equations:
\begin{equation} 
\label{eqn:QG_Model}
\begin{aligned}
 q_t &= \psi_x - \varepsilon J(\psi, q) - A \Delta^3 \psi + 2 \pi \sin(2 \pi y) \,, \\
   q & = \Delta \psi - F \psi \,,  \\
J(\psi, q) & \equiv \psi_x q_x - \psi_y q_y \,,
\end{aligned}
\end{equation}
where $\Delta := \partial^2/\partial x^2 + \partial^2/\partial y^2$ and $\psi$ is either the stream function or the surface elevation. 
We use the values of the model coefficients ~\eqref{eqn:QG_Model}  from \cite{sakov2008deterministic}, as follows: 
 $F=1600$, $\varepsilon=10^{-5}$, and $A=2\times 10^{-12}$.
The domain of the model is a $1 \times 1$ [space units] square, with $0\leq x \leq 1,\, 0\leq y \leq 1 \,,$  and is discretized by a grid of size $129 \times 129$ (including boundaries). Boundary conditions used are 
$\psi = \Delta \psi = \Delta^2 \psi = 0$. 
The model state dimension  is $\nvar = 16641$, while the model trajectories belong to affine subspaces with dimensions of the order of $10^2 - 10^3$~\cite{sakov2008deterministic}.

The time integration scheme used is the fourth-order Runge-Kutta scheme with a time step $1.25$ [time units].
%

% --- This is no longer the case; We now read the ensemble from the repository provided originally in the enkf-matlab package
% So basically, we are doing what the refernce papers are doing, but they don't mention how they generate the initial ensemble. 
%     We will follow theier steps in that as well.
% An initial background state is created by adding Gaussian noise with zero mean and variance equal to $5$ [units squared] to the reference initial condition . 
% An initial ensemble is created by adding Gaussian random perturbations with zero mean and variance $5$ [units squared]  to the forecast initial condition.

For all experiments in this work, the model is run over $1000$ model time steps, with observations made available every $10$ time steps.
In this synthetic model the scales are not relevant, and we use generic space, time, and solution amplitude units.

%~~~~~~~~~~~~~~~~~~~~~~~~~~~~~~~~~~~~~~~~~~~~~~~~~~~~~~~~~~~~~~~~~~~
\subsubsection{Observations and observation operators} % \label{subsec:QG_Observations}
%~~~~~~~~~~~~~~~~~~~~~~~~~~~~~~~~~~~~~~~~~~~~~~~~~~~~~~~~~~~~~~~~~~~
%
Two observation operators are used with this model. 
\begin{itemize}
\item First we use a standard linear operator to observe $300$ components of $\psi$. The observation 
error variance is $4.0$ ~[units squared]. Synthetic the observations are obtained by adding white noise to measurements of the see height level (SSH) extracted from a model 
run with lower viscosity.  
\item The second observation operator measures the magnitude of the flow velocity $\sqrt{u^2 + v^2}$.
The flow velocity components $u,\,v $ are obtained using a finite difference approximation of the following relations to the stream function:
\begin{equation}
\label{eqn:qg_wind_velocity}
  u = + \frac{\partial \psi}{\partial y}\,, \quad v = - \frac{\partial \psi}{\partial x}.
\end{equation}
\end{itemize}
In both cases, the observed components are uniformly distributed over the state vector length, with a random offset, that is updated at each assimilation cycle.

\begin{figure}[!htbp]
\centering
\subfigure[Reference initial state and observational grid]{
  \includegraphics[width=0.40\linewidth]{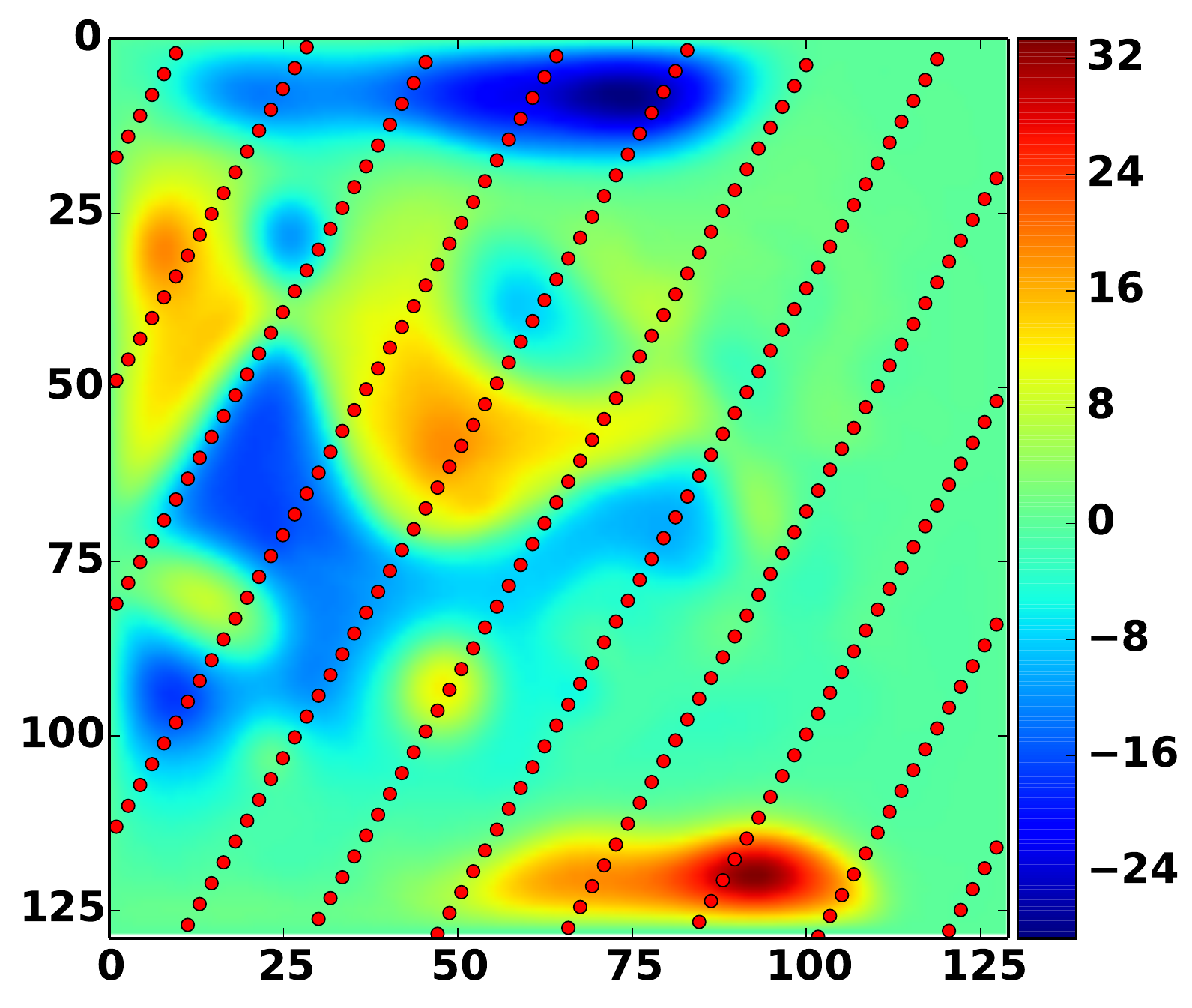}
  \label{fig:QG1p5_Ref_IC_with_Obs_Grid}
}
\quad % \hfill
\subfigure[Mean of the initial forecast ensemble]{
  \includegraphics[width=0.40\linewidth]{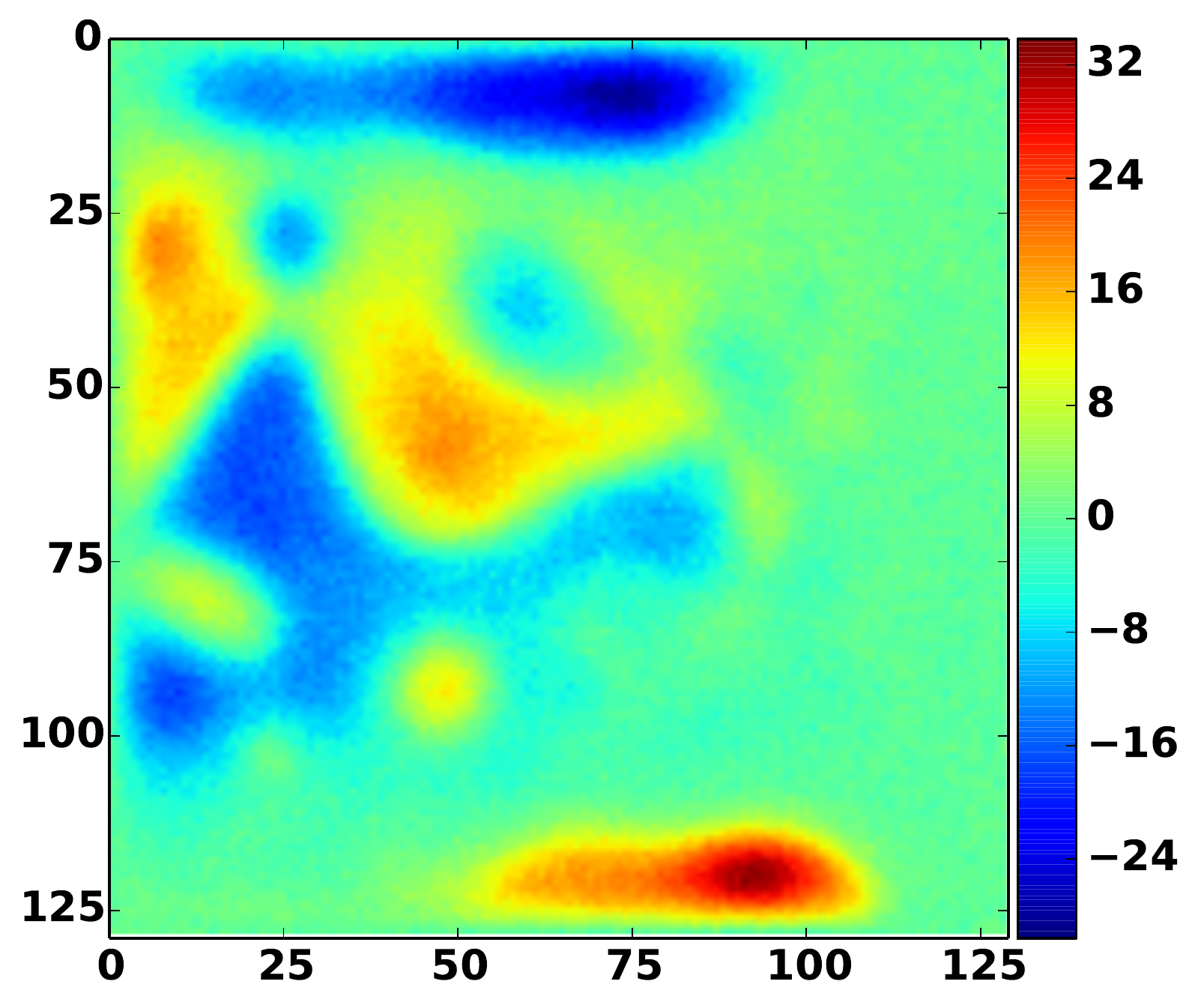}
  \label{fig:QG1p5_Frcst_IC}
}
\caption{The QG-1.5 model.  The red dots in panel \ref{fig:QG1p5_Ref_IC_with_Obs_Grid} indicate the location of observations for one of the test cases employed.
 }
  \label{fig:QG1p5_Initial_Reference_and_Forecast}
\end{figure}

The reference initial state along with an example of the observational grid used, and the initial forecast state are shown in 
Figure~\ref{fig:QG1p5_Initial_Reference_and_Forecast}.
%

%~~~~~~~~~~~~~~~~~~~~~~~~~~~~~~~~~~~~~~~~~~~~~~~~~~~~~~~~~~~~~~~~~~~
\subsubsection{Filter tuning}
% \label{Subsec:QG_Results}
%~~~~~~~~~~~~~~~~~~~~~~~~~~~~~~~~~~~~~~~~~~~~~~~~~~~~~~~~~~~~~~~~~~~
%
We used a deterministic implementation of EnKF (DEnKF) with parameters tuned as suggested in~\cite{sakov2008deterministic}. 
Specifically, we apply a covariance localization by means of a Hadamard product as explained in~\cite{Houtekamer_2001}.
The localization function used is Gaspari-Cohn~\cite{Gaspari_1999_correlation} with localization radius set to $12$ grid cells.
Inflation is applied with factor $\delta=1.06$ to the analysis ensemble of anomalies at the end of each assimilation cycle of DEnKF.

The parameters of the HMC and \ClHMC sampling filters are tuned empirically in a preprocessing step in the HMC filter to guarantee a rejection rate 
at most between $25\%$ to $30\%$.
Here we tune the parameters of the Hamiltonian trajectory only once at the beginning of the assimilation experiment.
Specifically, the step size parameters of the symplectic integrator are set to $h=0.075,\,m=25$ in the presence of the linear observation operator, 
and are set to $h=0.015,\,m=25$ when the nonlinear observation operator~\eqref{eqn:qg_wind_velocity} is used. % Now the linear, and nonlinear cases are not the same 
The integrator used for the Hamiltonian system in all experiments  is the three-stage symplectic integrator~\cite{attia2015hmcfilter,sanz2014Markov}.
The mass matrix $\mathbf{M}$ is chosen to be a diagonal matrix whose diagonal is set to the diagonal of the precision matrix of the forecast ensemble.
In the current experiments, the first $50$ steps of the Markov chains are discarded as a burn-in stage. Alternatively, one can run a suboptimal minimization of the negative-log of the posterior to achieve convergence to the posterior.

The parameters of the MC-\ClHMC filter are set as follows. 
The step size parameters of the symplectic integrator are set to $h=0.05/\nc$, $m=15$ in the experiments with linear observation operator, and $h=0.0075/\nc$, $m=15$  in the case 
of the nonlinear observation operator~\eqref{eqn:qg_wind_velocity}.
The mass matrix is a diagonal matrix whose diagonal is set to the diagonal of the precision matrix of the forecast ensemble labeled under the corresponding mixture component. 
To avoid numerical problems related to very small ensemble variances, for example in the case of outliers, the variances are averaged with the modeled forecast variances of $5$ [units squared]. 

The prior GMM is built with number of components determined using AIC model selection criteria, with a lower bound of $5$ of the number of ensemble members belonging to each component of the mixture. 
This lower bound is enforced as a means to ameliorate the effect of outliers on the GMM construction.
%
% Results for both the \ClHMC and the MC-\ClHMC filters are shown for the number of mixture components detected using AIC criteria. 
%
In all experiments involving \ClHMC, and MC-\ClHMC, the diagonal covariances relaxation assumption is imposed. However, this structure is not imposed if only one mixture component is detected, and \ClHMC and MC-\ClHMC filters fall back to the original HMC filter.
% The next statement, can be moved to conclusion!
For cases where a component contains a very small number of ensemble members covariance tapering~\cite{furrer2006covariance} can prove useful.

%~~~~~~~~~~~~~~~~~~~~~~~~~~~~~~~~~~~~~~~~~~~~~~~~~~~~~~~~~~~~~~~~~~~
\subsubsection{Assessment metrics} 
%~~~~~~~~~~~~~~~~~~~~~~~~~~~~~~~~~~~~~~~~~~~~~~~~~~~~~~~~~~~~~~~~~~~
%
To assess the accuracy of the tested filters we use the root mean squared error (RMSE):
\begin{equation} 
\label{eqn:RMSE}
 {\rm RMSE} = \sqrt{\frac{1}{\nvar} \sum_{i=1}^{\nvar}{(x_i - x_i^{\rm true})^2} } \, ,
\end{equation}
where $\x^{\rm true}={\psi}^{\rm true}$ is the reference state of the system and $\x$ is the analysis state, e.g. the average of the analysis ensemble. 
Here $\nvar=129 \times 129=16641$ is the dimension of the model state.We also use Talagrand (rank) histogram~\cite{anderson1996method,candille2005evaluation} to assess the quality of the ensemble spread around the true state.

%~~~~~~~~~~~~~~~~~~~~~~~~~~~~~~~~~~~~~~~~~~~~~~~~~~~~~~~~~~~~~~~~~~~
\subsubsection{Results with linear observation operator} 
\begin{figure}[!htbp]
  \centering
  \includegraphics[width=0.75\linewidth]{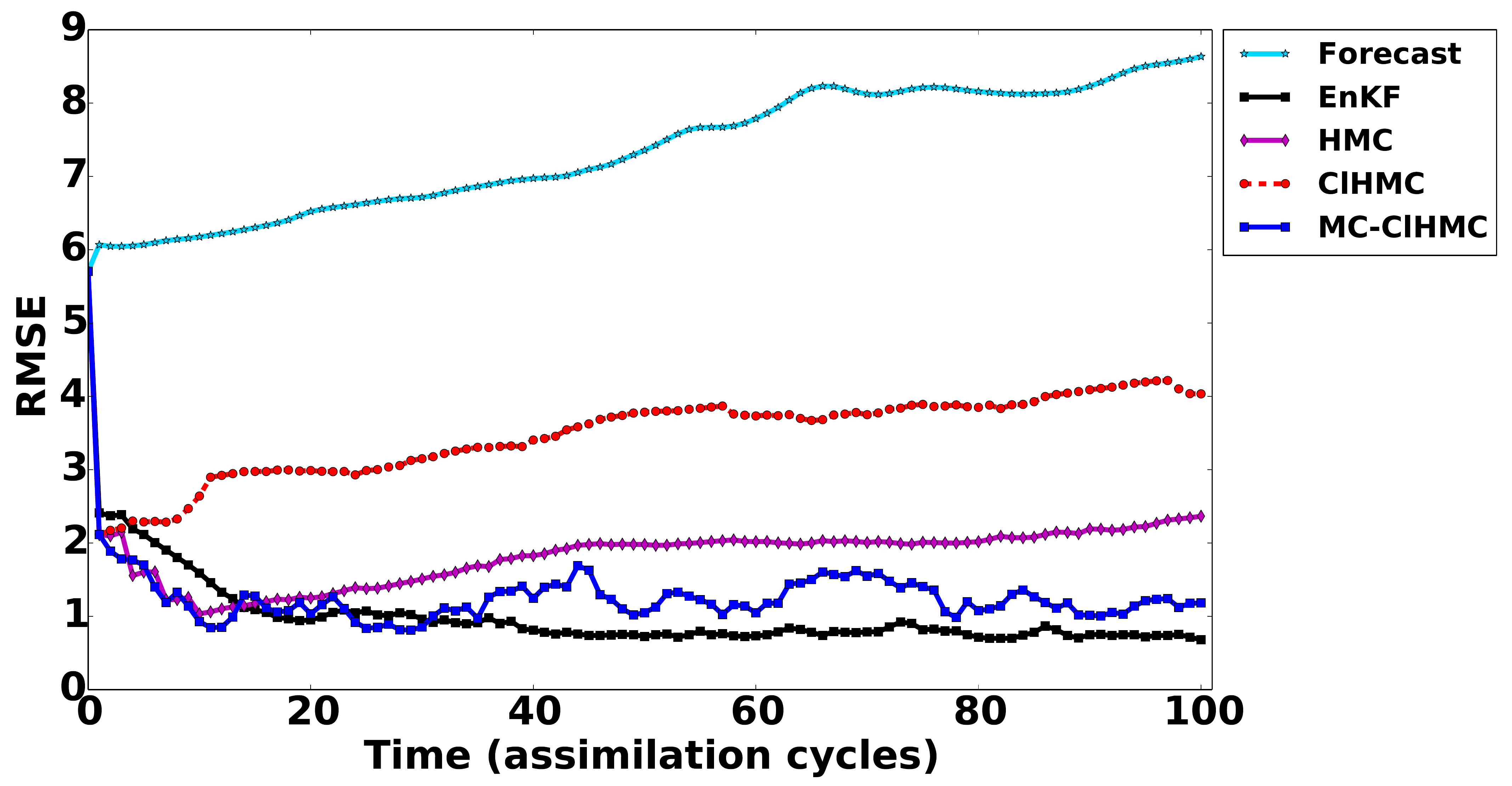}
  \caption{Data assimilation results with the linear observation operator.   
   RMSE of the \eqref{eqn:RMSE} analyses obtained by EnKF, HMC, \ClHMC,  and MC-\ClHMC filters.}
  \label{fig:QG_linear_H_RMSE}
\end{figure}

%~~~~~~~~~~~~~~~~~~~~~~~~~~~~~~~~~~~~~~~~~~~~~~~~~~~~~~~~~~~~~~~~~~~
%
Figure~\ref{fig:QG_linear_H_RMSE} presents the RMSE \eqref{eqn:RMSE} results of the analyses obtained using EnKF, HMC, \ClHMC , and MC-\ClHMC filters in the presence of a linear observation operator. 
Figure~\ref{fig:QG_linear_H_RMSE} shows that the results of all HMC filter versions improve quickly at the first few assimilation windows.
While the results of the original HMC filter improve quickly at the first few assimilation windows, the performance of the original HMC filter degrades compared to the DEnKF filter performance especially in the long run.
We believe that the two main factors contribute to the HMC filter degradation are the parameter tuning, and the development of non-Gaussianity in the prior distribution.
The \ClHMC analysis drifts away quickly from the true trajectory. This is mainly because the HMC sampling strategy is unable to cover all probability modes in the posterior distribution.
To guarantee that the sampling filter covers the truth well, the sampler has to be able to sample properly from all posterior probability modes.
This is achieved by design by the MC-\ClHMC filter. The MC-\ClHMC version produces RMSE results comparable to the RMSE obtained by DEnKF.

% As seen in Figure~\ref{fig:QG_linear_H_RMSE}, the \ClHMC analysis drifts away from the true trajectory in the long run.
% The original HMC filter (without optimal tuning of the parameter set) produces an analysis that is close to the EnKF analysis as measured by RMSE.
%
%
As discussed in~\cite{attia2015hmcfilter} the performance of HMC filter can be further enhanced by automatically tuning the parameters of the symplectic  integrator at the beginning of 
each assimilation cycle. Here however we are mainly interested in assessing the performance of the new methodologies compared to the original HMC filter 
using equivalent settings. 

It is important to note that the MC-\ClHMC filter requires shorter Hamiltonian trajectories to explore the space under each local mixture component, which results in computational savings. 
Additional savings can be obtains by running the chains in parallel to sample different regions of the posterior.

\begin{figure}[!htbp]
  \center
  \subfigure[DEnKF]{
    \includegraphics[width=0.45\linewidth]{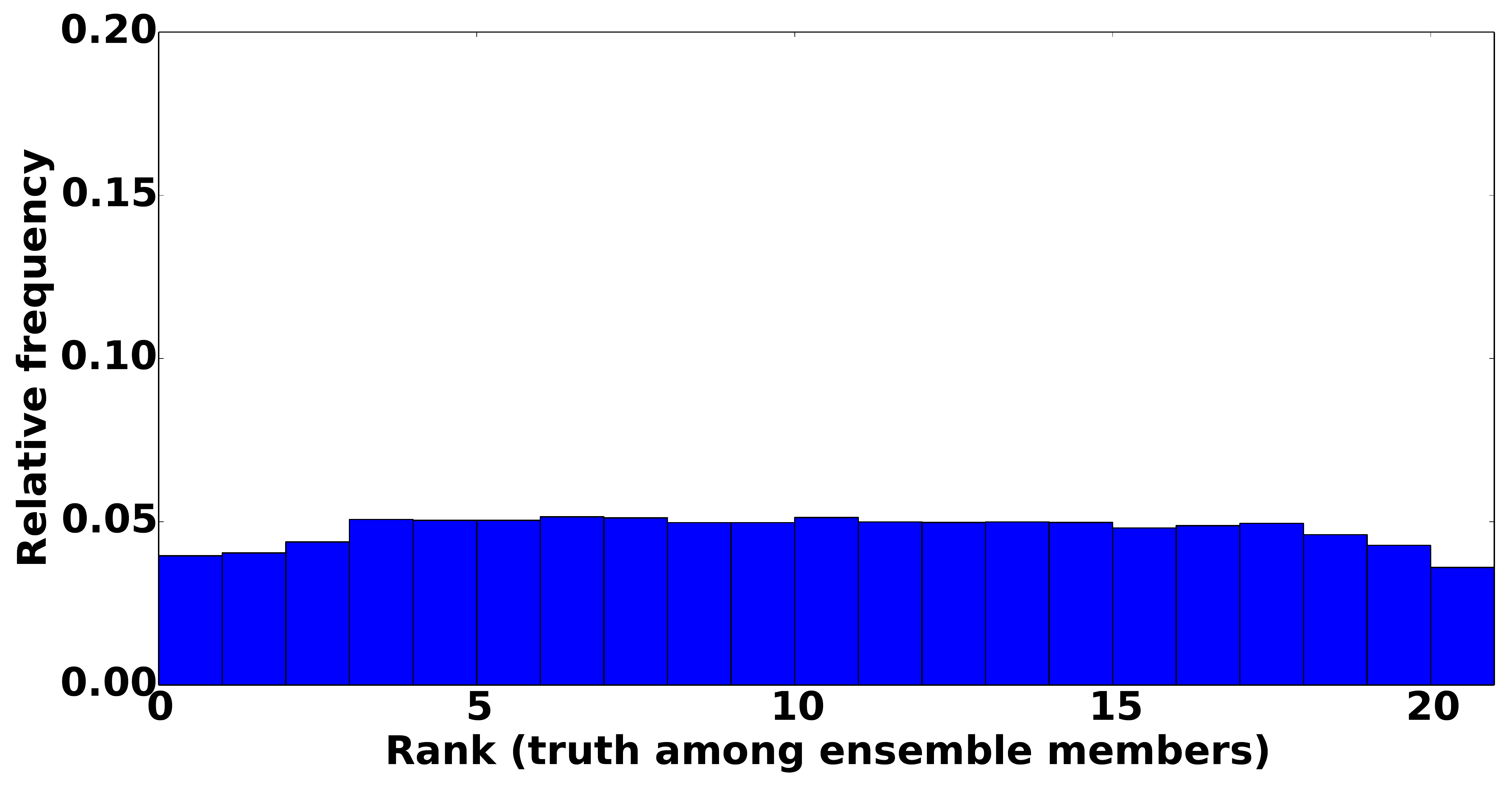}
    \label{fig:QG_linearH_EnKF_RankHist}
  }
  \quad % \hfill  
  \subfigure[HMC]{
    \includegraphics[width=0.45\linewidth]{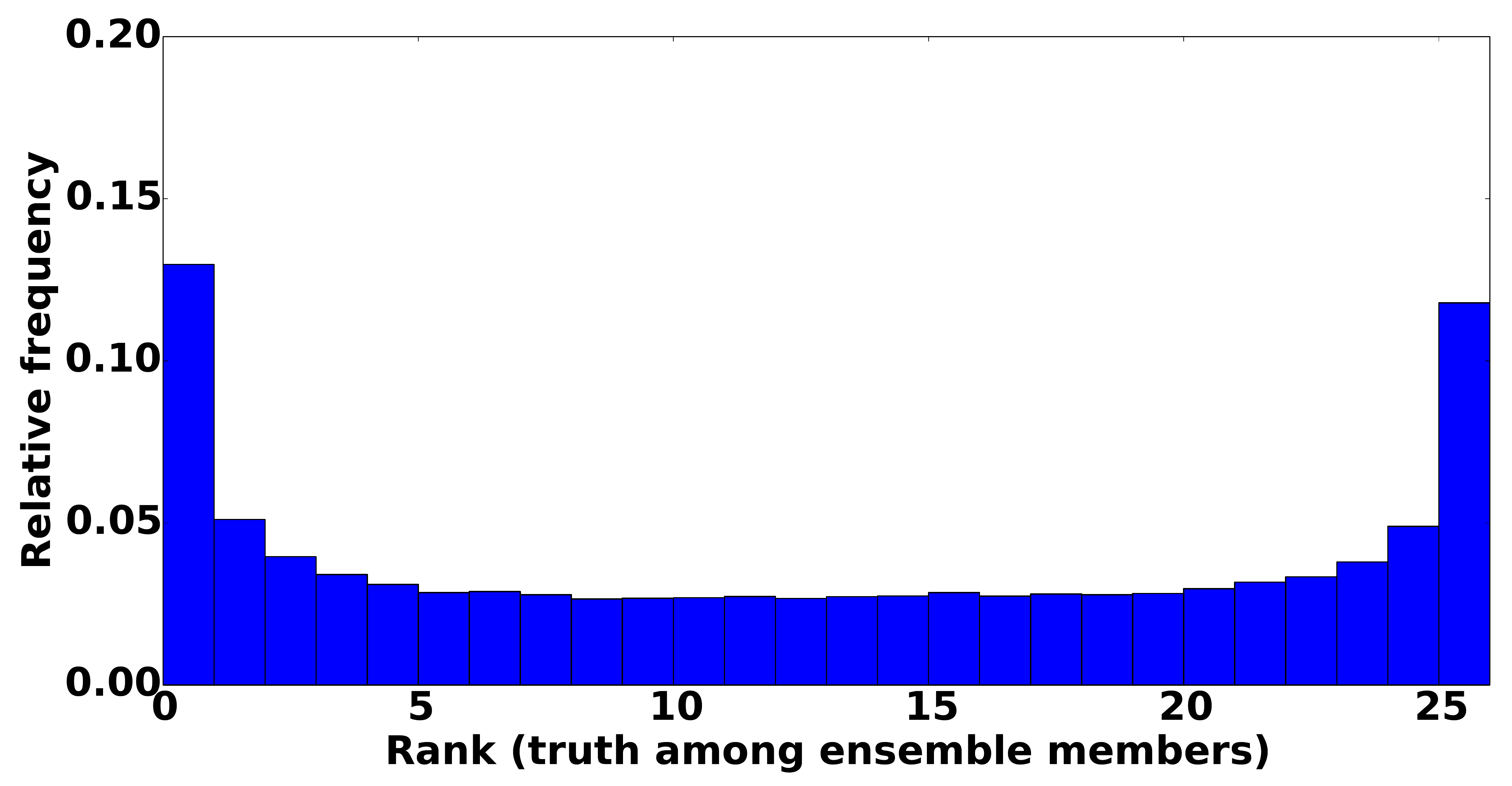}
    \label{fig:QG_linearH_HMC_RankHist}
  }
  \\ %\hfill
  \subfigure[\ClHMC+AIC]{
    \includegraphics[width=0.45\linewidth]{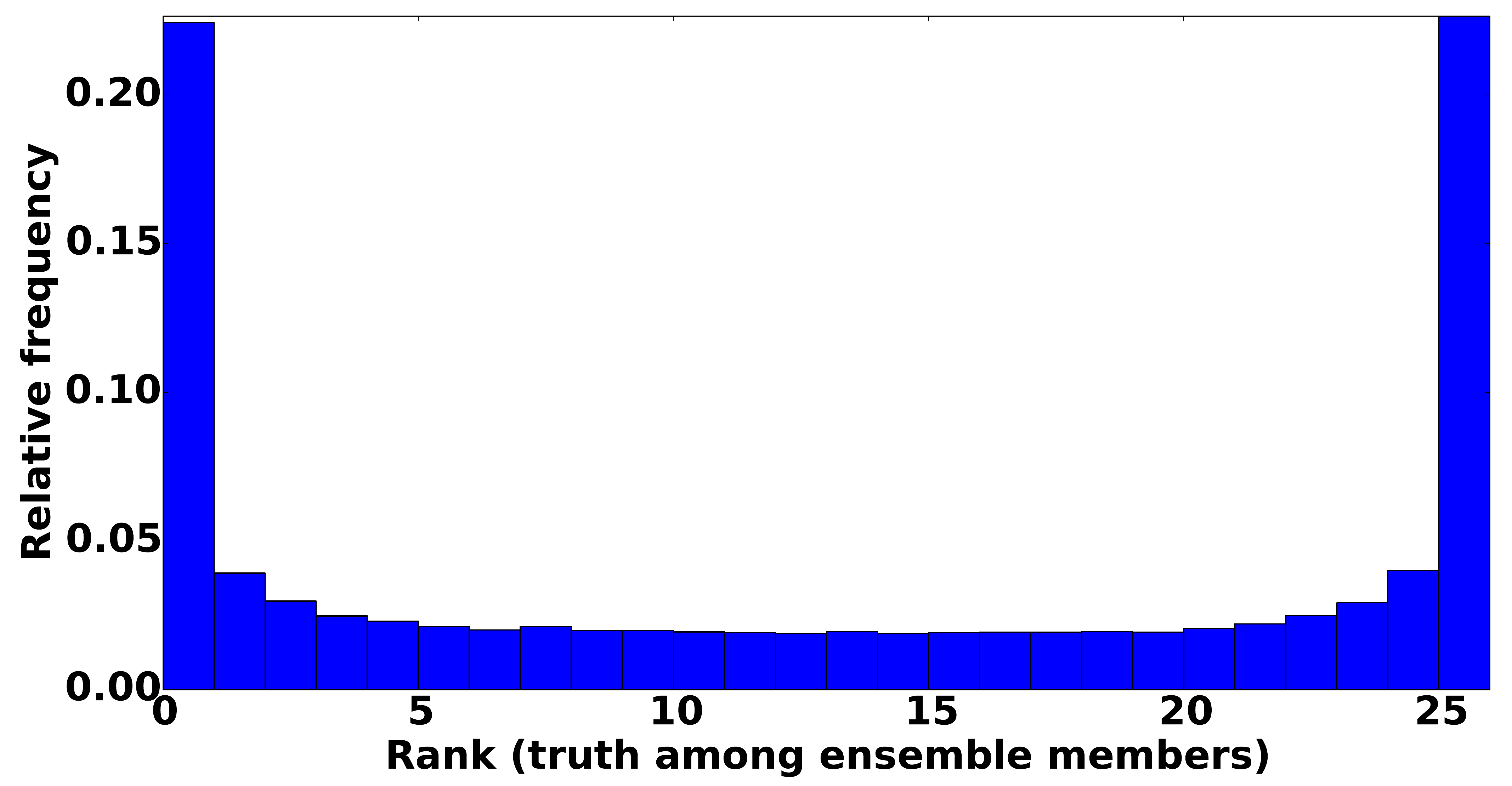}
    \label{fig:QG_linearH_ClHMC_AIC_RankHist}
  }
  \quad % \hfill
  \subfigure[MC-\ClHMC+AIC]{
    \includegraphics[width=0.45\linewidth]{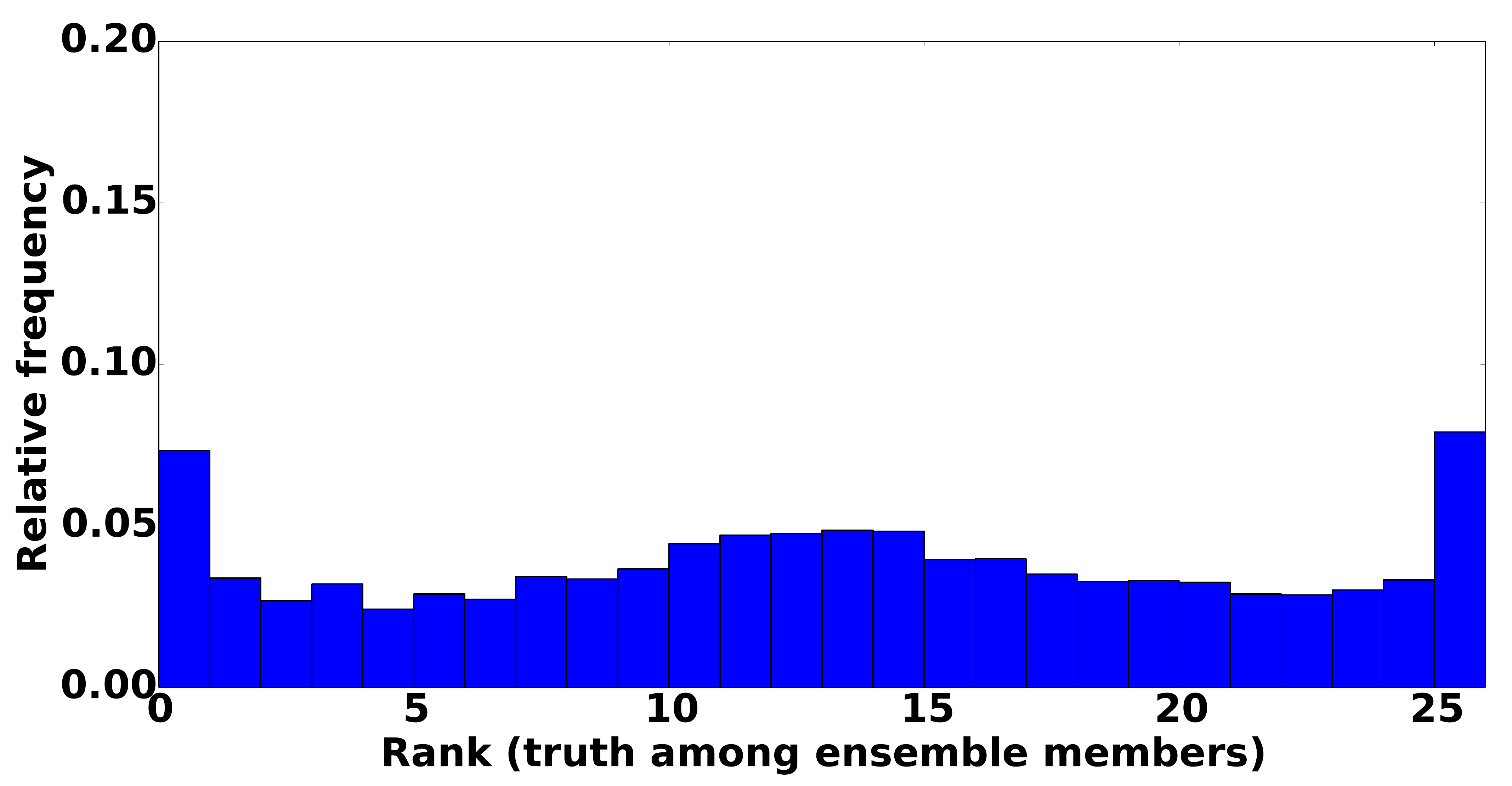}
    \label{fig:QG_linearH_MC_ClHMC_AIC_RankHist}
  }
  \caption{Data assimilation results with the linear observation operator. 
	   The rank histograms of where the truth ranks among posterior ensemble members. 
	   The ranks are evaluated for every $16^{th}$ variable in the state vector (past the correlation bound) at $100$ assimilation times.
	  }
  \label{fig:QG_linear_H_RankHist}
\end{figure}

Since we are not interested in only a \textit{single} best estimate of the true state of the system, RMSE alone is not sufficient to judge the quality of the filtering system. 
The analysis ensemble sampled from the posterior should be spread widely enough to cover the truth and avoid filter collapse. 
The rank histograms of the analysis ensembles are shown in Figure~\ref{fig:QG_linear_H_RankHist}. 
The two small spikes in Figure~\ref{fig:QG_linearH_HMC_RankHist} suggest that the performance of the original HMC filter could be enhanced by increasing the length of the Hamiltonian trajectories 
in some assimilation cycles.

The rank histogram shown in Figure~\ref{fig:QG_linearH_ClHMC_AIC_RankHist} shows that the analysis ensembles produced by the \ClHMC filter tend to be under-dispersed.
Since the ensemble size is relatively small and the prior GMM is multimodal, with regions of low-probability between the different mixture components, 
a multimodal mixture posterior with isolated components is obtained. As explained in~\cite{nishimura2016geometrically}, 
this is a case where HMC sampling in general can suffer from being entrapped in a local minimum (and fails to jump between different high probability modes). 
This behavior is expected to result in ensemble collapse, as seen in Figure~\ref{fig:QG_linearH_ClHMC_AIC_RankHist}, 
leading to filter degradation in the long run as illustrated by the RMS errors shown in Figure~\ref{fig:QG_linear_H_RMSE}.

\begin{figure}[!htbp]
  \center
  \quad % \hfill  
  \subfigure[\ClHMC+AIC; first two observation windows]{
    \includegraphics[width=0.45\linewidth]{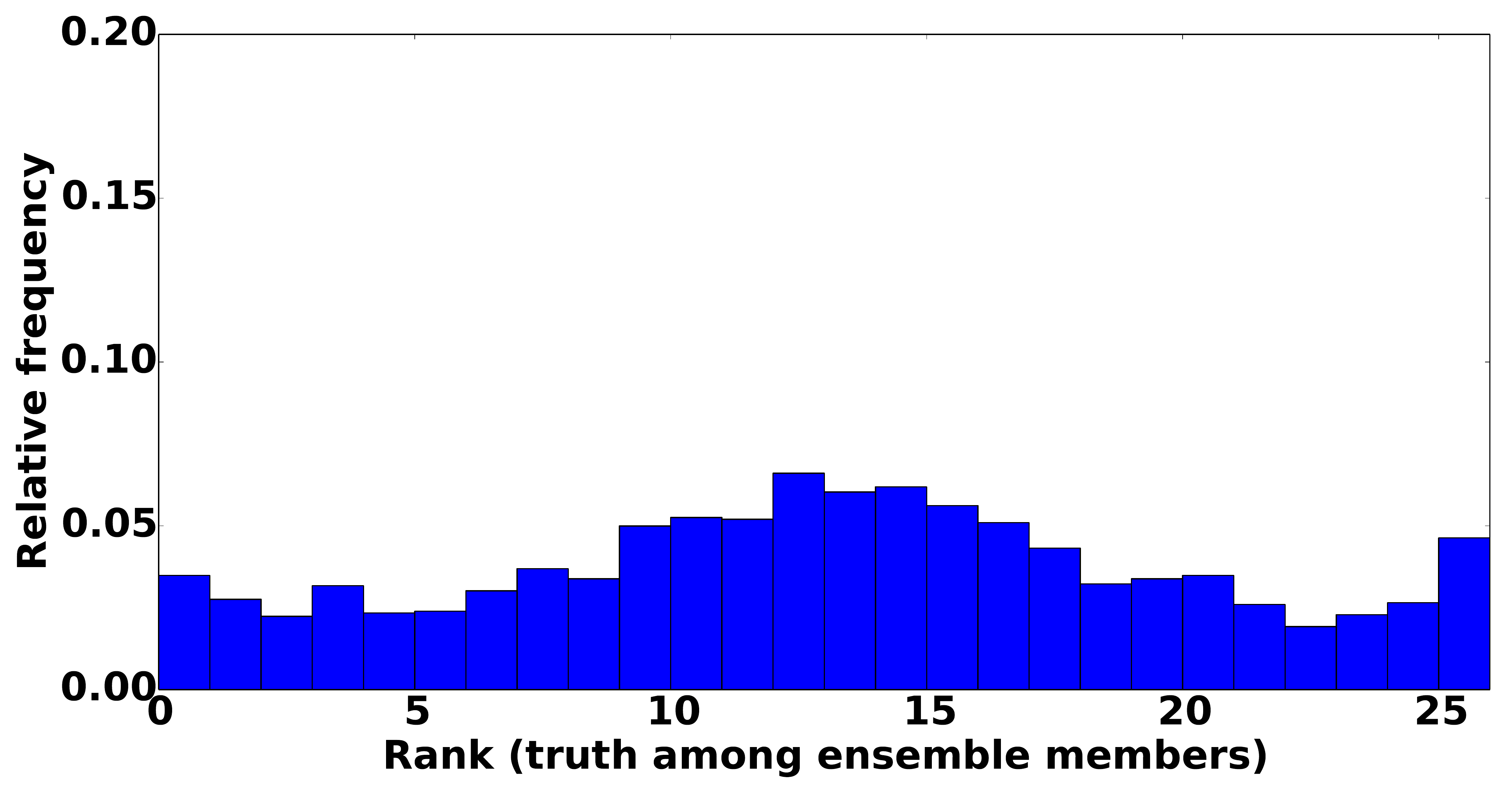}
    \label{fig:QG_linearH_ClHMC_AIC_RankHist_2obs_Only}
  }
  \\ %\hfill
  \subfigure[\ClHMC+AIC; first five observation windows]{
    \includegraphics[width=0.45\linewidth]{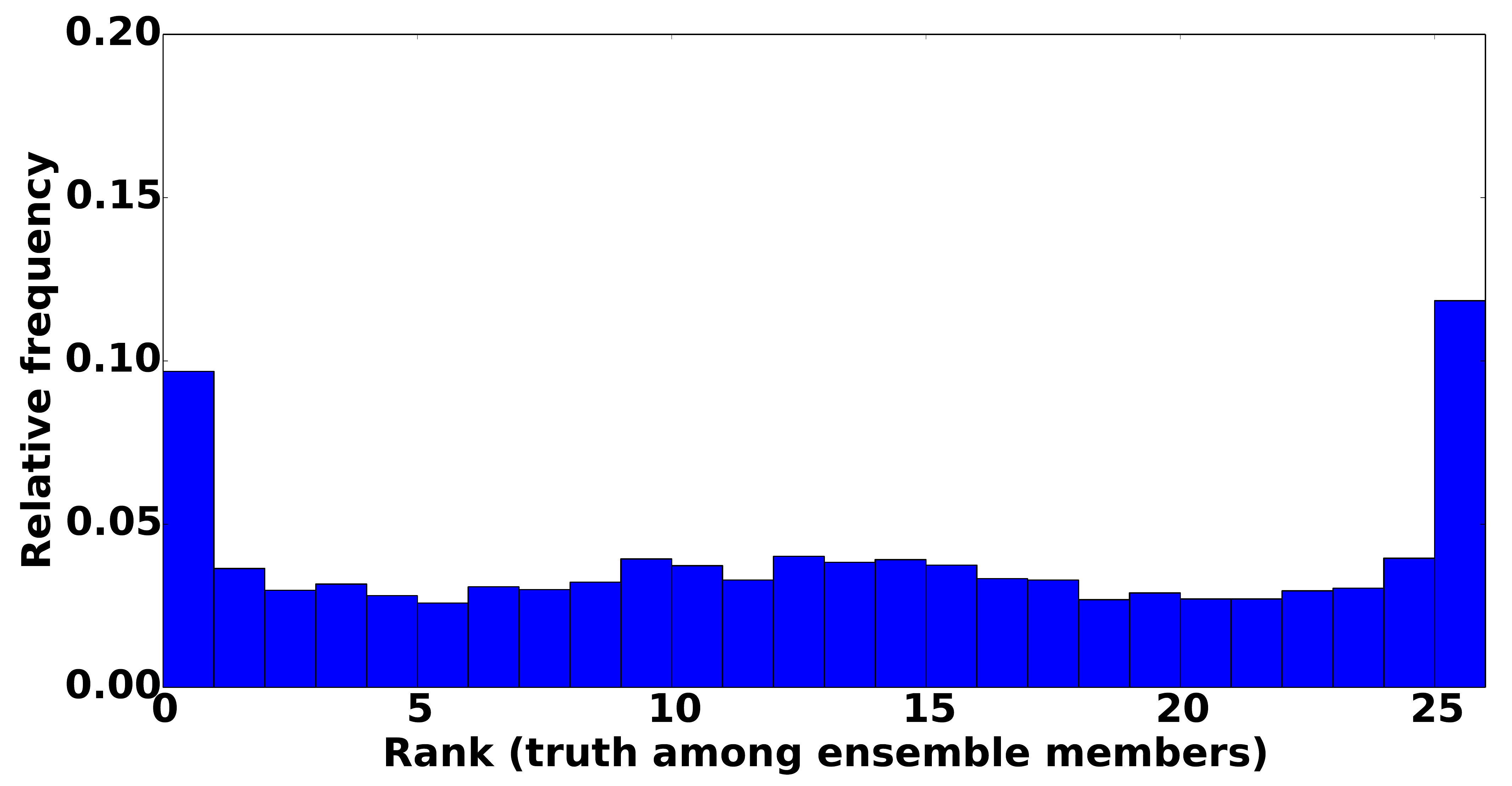}
    \label{fig:QG_linearH_ClHMC_AIC_RankHist_5obs_Only}
  }
  \quad % \hfill
  \subfigure[\ClHMC+AIC; first ten observation windows]{
    \includegraphics[width=0.45\linewidth]{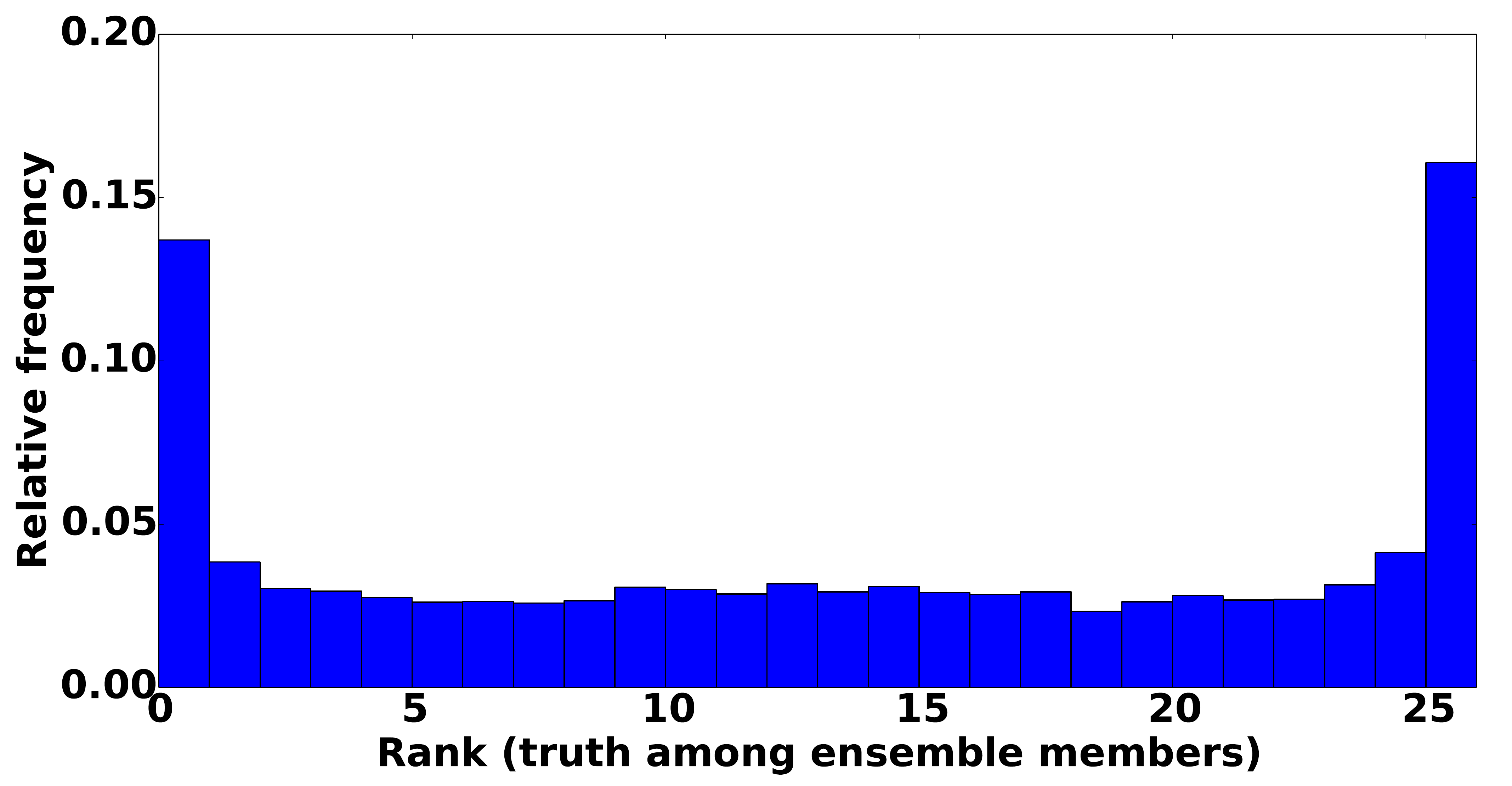}
    \label{fig:QG_linearH_ClHMC_AIC_RankHist_10obs_Only}
  }
  \caption{Data assimilation results using a linear observation operator.
   Rank histograms of where the truth ranks among posterior ensemble members. The ranks are evaluated for every $16^{th}$ variable in the state vector (past the correlation bound).
   Rank histograms of \ClHMC  results obtained at the first two, five, and ten assimilation cycles respectively are shown.
   The model selection criterion used is AIC.
   }
  \label{fig:QG_linearH_ClHMC_minimal_RankHist}
\end{figure}

The results shown in Figure~\ref{fig:QG_linearH_ClHMC_AIC_RankHist} suggest that the analysis ensemble collected by \ClHMC fails to cover all mixture components, 
thereby losing its dispersion when it is applied repetitively.
This is supported by the results in Figure~\ref{fig:QG_linearH_ClHMC_minimal_RankHist}, where the rank histograms are plotted using results from the first two, five, and ten cycles, respectively.

The ensemble collapse can be avoided if we force the sampler to collect ensemble members from all the probability modes.  
This is illustrated by the rank histograms of results obtained using the MC-\ClHMC filter with AIC criteria as shown in~Figure~\ref{fig:QG_linearH_MC_ClHMC_AIC_RankHist}.

We believe that having isolated regions of high probability, e.g., with very small number of ensemble members in each component, 
can be the critical factor leading the poor long-term performance of \ClHMC. 
This is alleviated here by imposing a minimum number of $3$ ensemble points in each component, e.g. via hard assignment, of the mixture while constructing the GMM approximation of the prior.

With automatic tuning of the Hamiltonian parameters the performance of both HMC and MC-\ClHMC filters is expected to be greatly enhanced.
We have only shown the results of \ClHMC, and MC-\ClHMC with AIC information criterion; experiments carried out using other model selection criteria such as BIC have proven to be very similar.

\begin{figure}[!htbp]
  \center
  \subfigure[EnKF; t=300]{
    \includegraphics[width=0.22\linewidth]{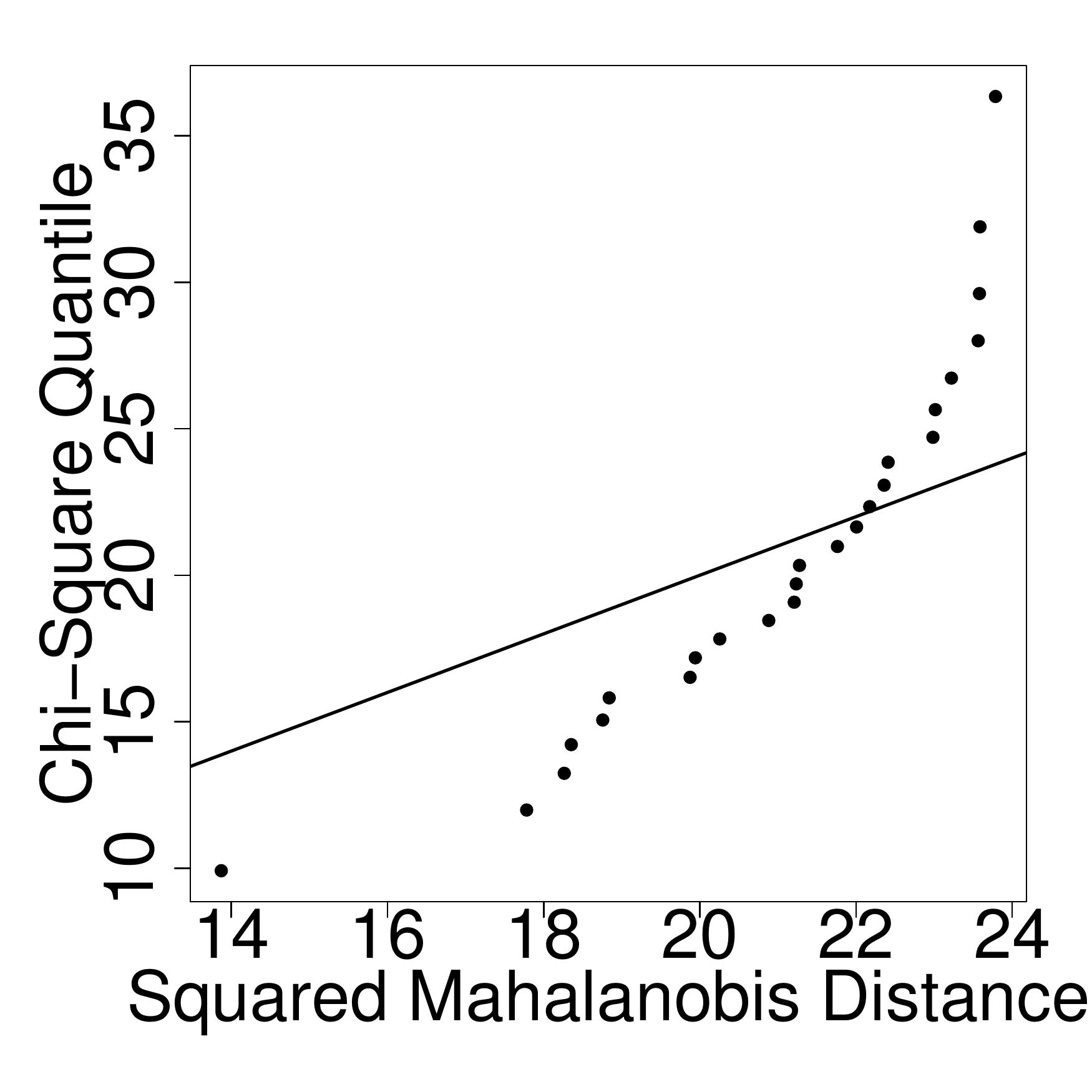}
    \label{fig:QG_qqplot_forecast_EnKF_cycle_23}
  }
  \subfigure[EnKF; t=775]{
    \includegraphics[width=0.22\linewidth]{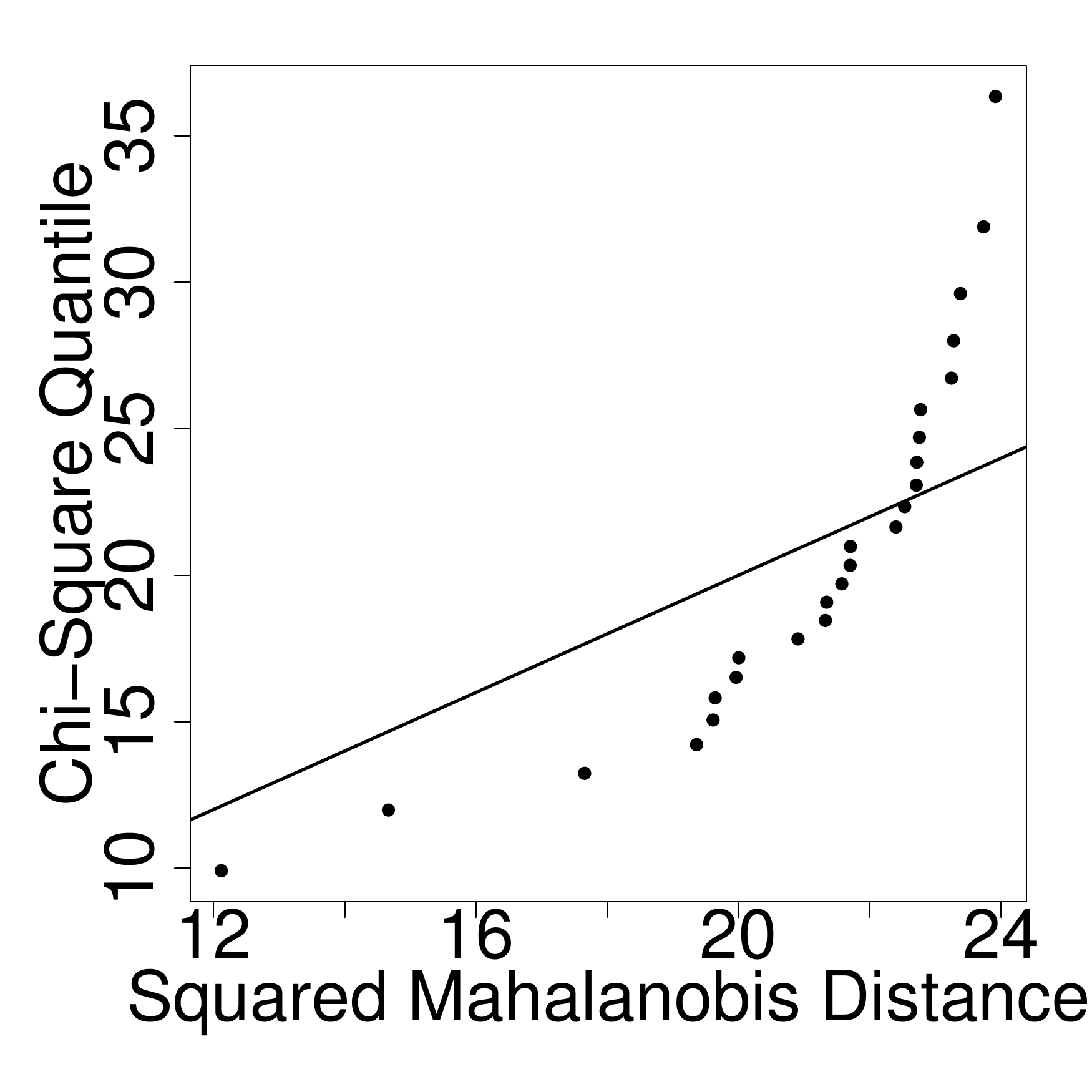}
    \label{fig:QG_qqplot_forecast_EnKF_cycle_61}
  }
  \subfigure[EnKF; t=1000]{
    \includegraphics[width=0.22\linewidth]{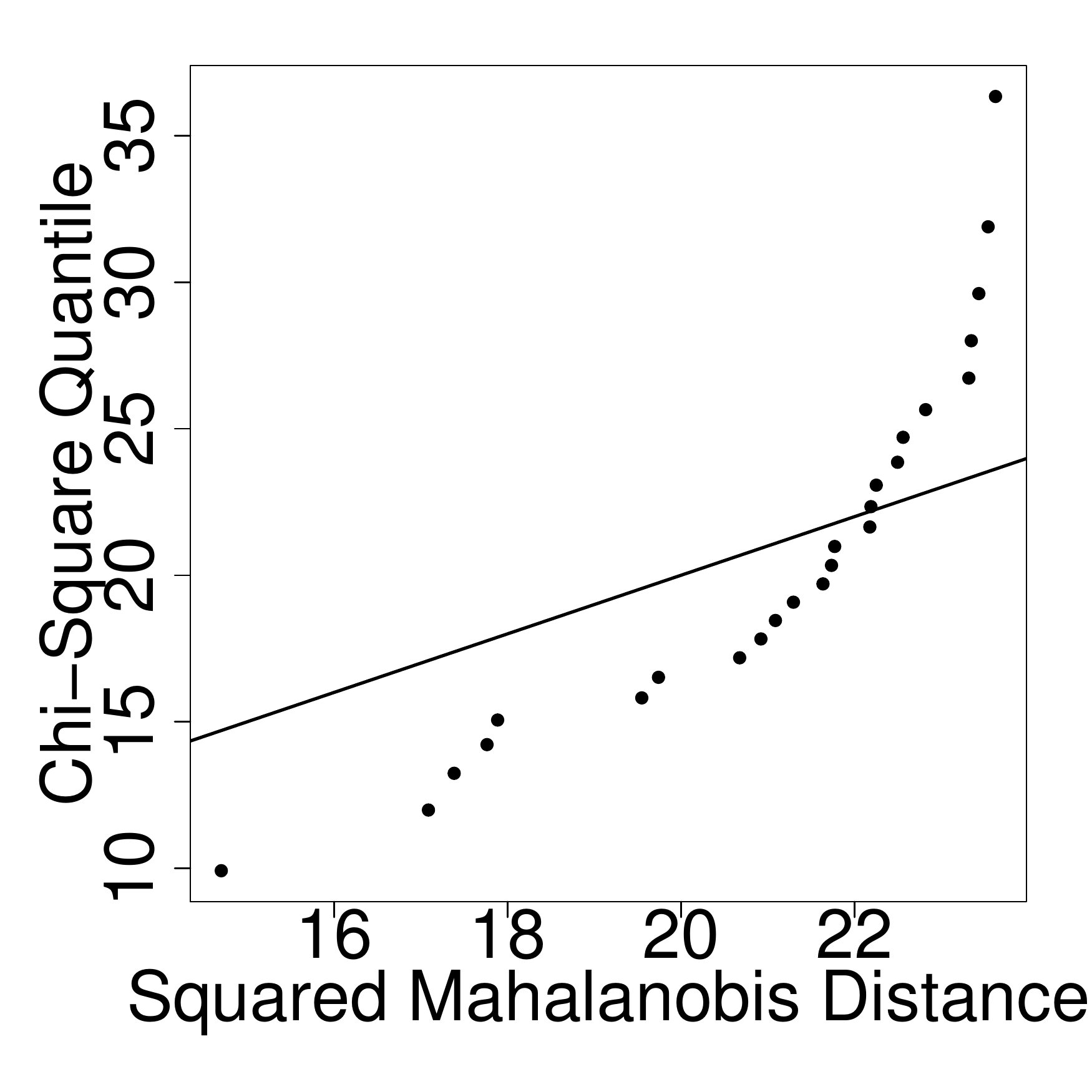}
    \label{fig:QG_qqplot_forecast_EnKF_cycle_80}
  }
  \subfigure[EnKF; t=1200]{
    \includegraphics[width=0.22\linewidth]{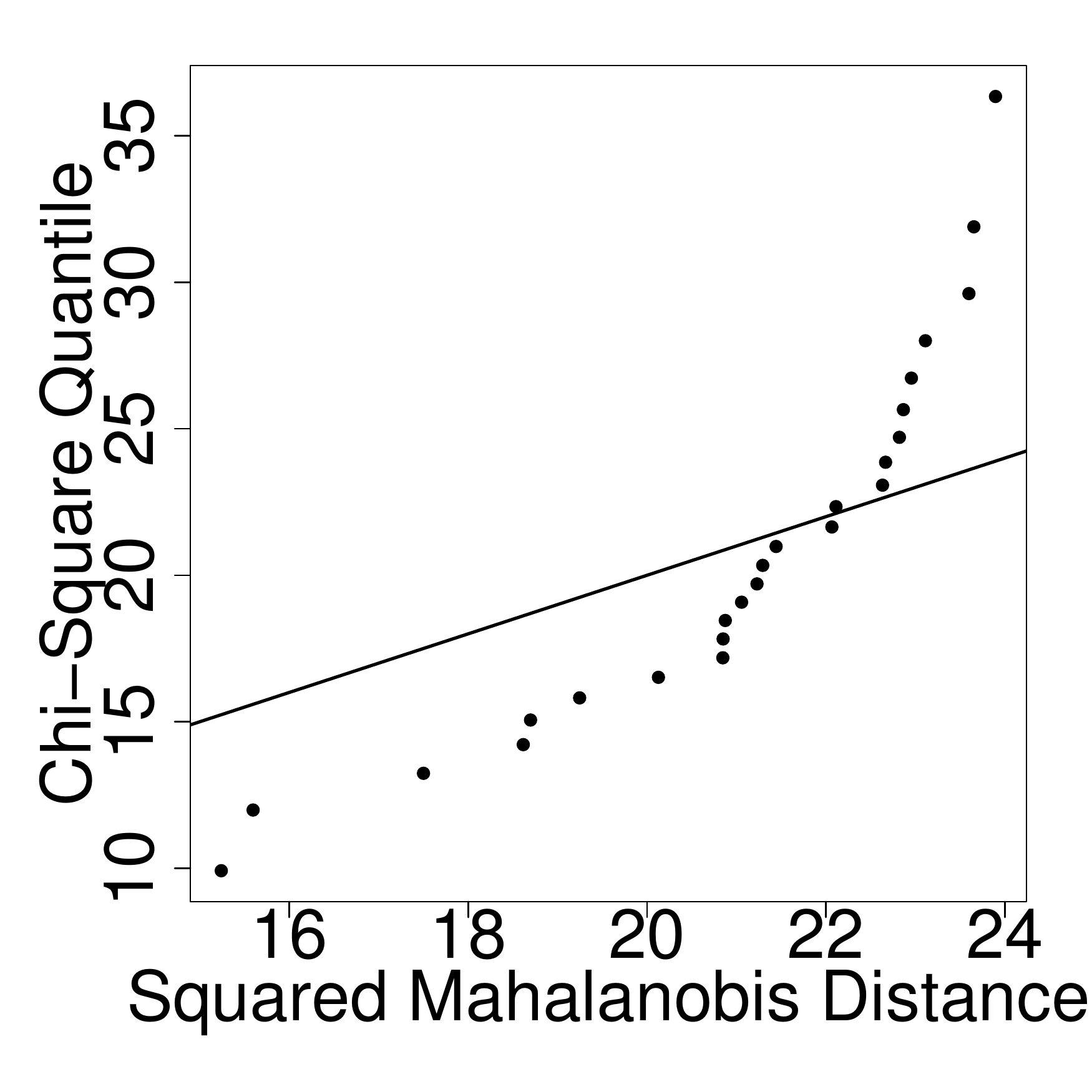}
    \label{fig:QG_qqplot_forecast_EnKF_cycle_8096}
  }
  \\
  \subfigure[HMC; t=300]{
    \includegraphics[width=0.22\linewidth]{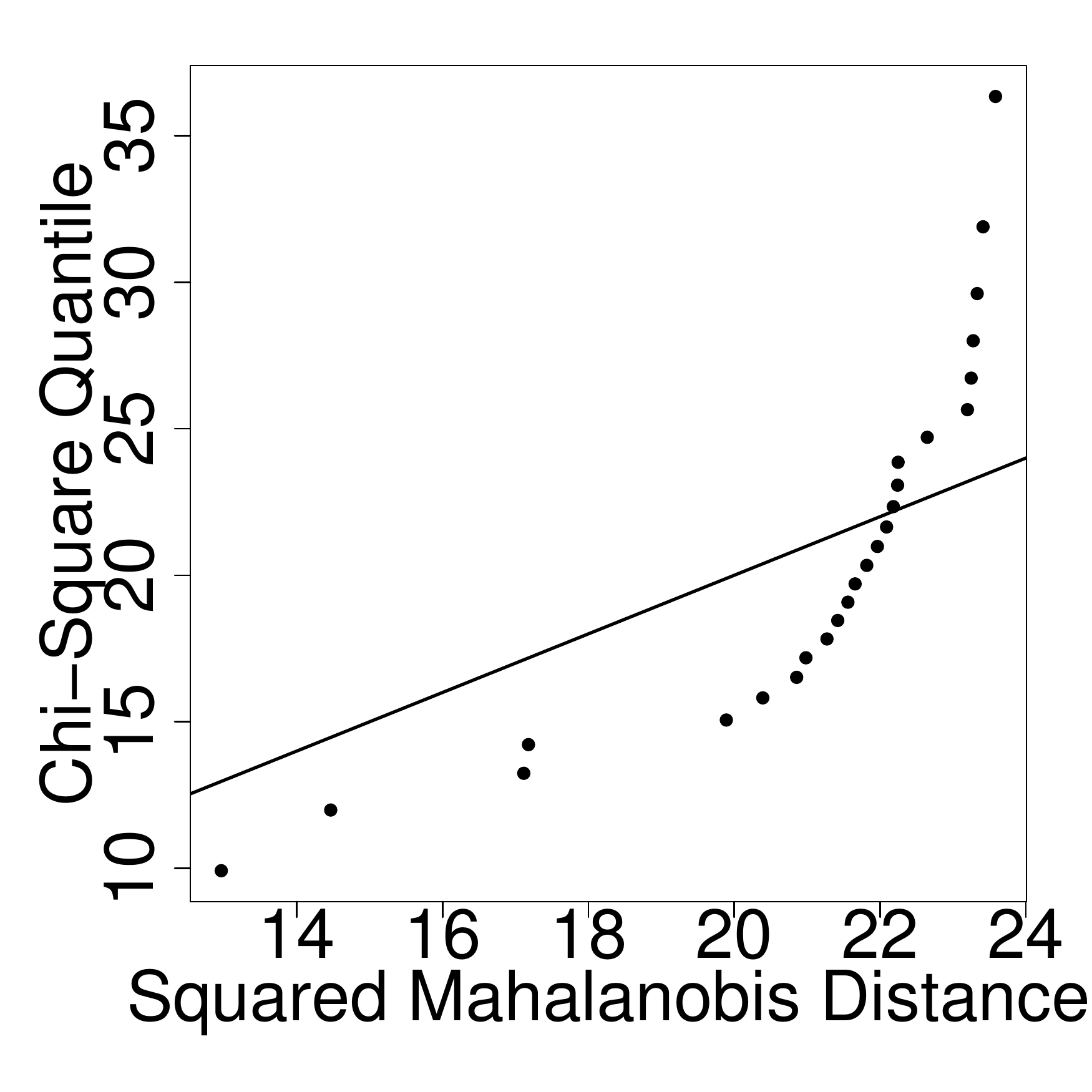}
    \label{fig:QG_qqplot_forecast_HMC_cycle_23}
  }
  \subfigure[HMC; t=775]{
    \includegraphics[width=0.22\linewidth]{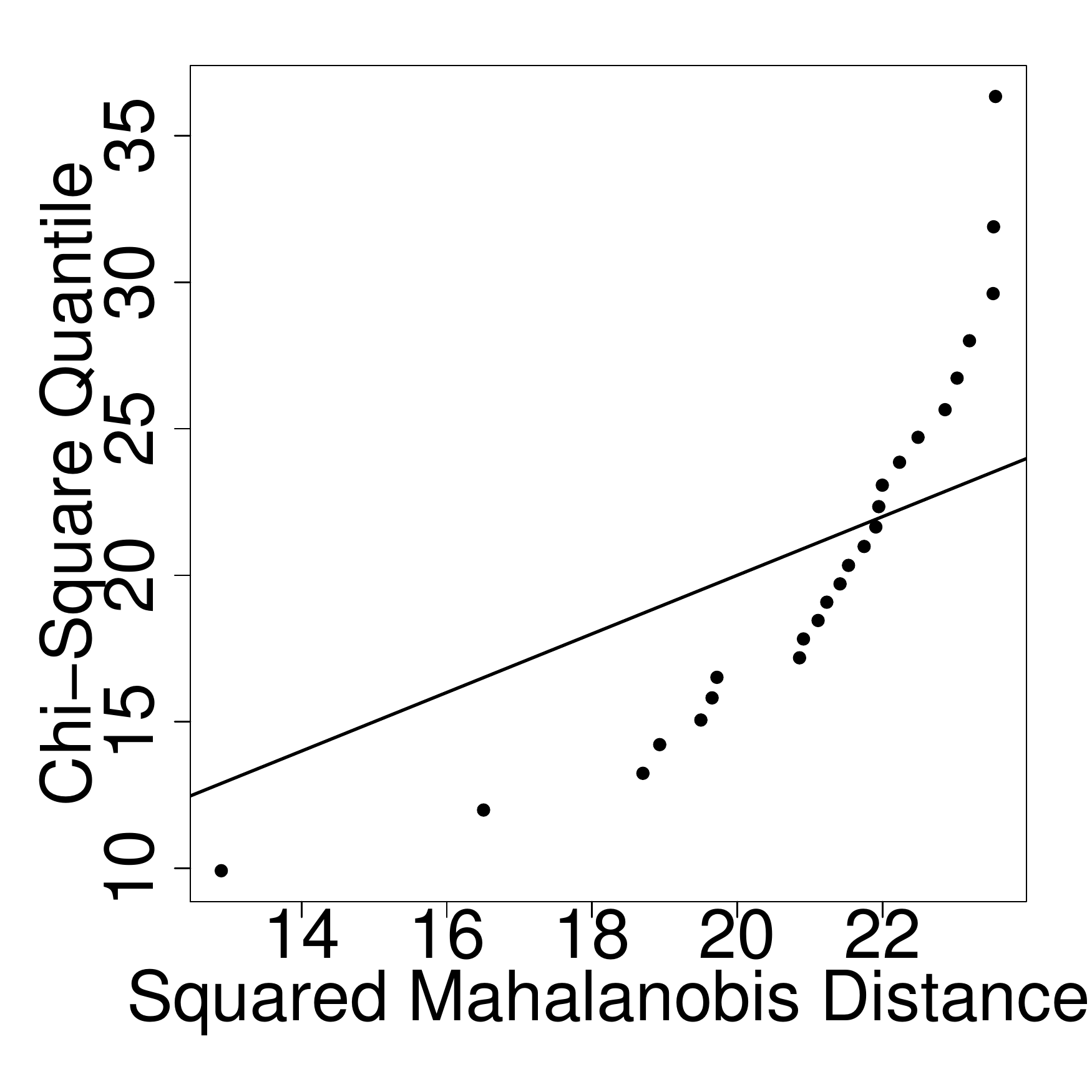}
    \label{fig:QG_qqplot_forecast_HMC_cycle_61}
  }
  \subfigure[HMC; t=1000]{
    \includegraphics[width=0.22\linewidth]{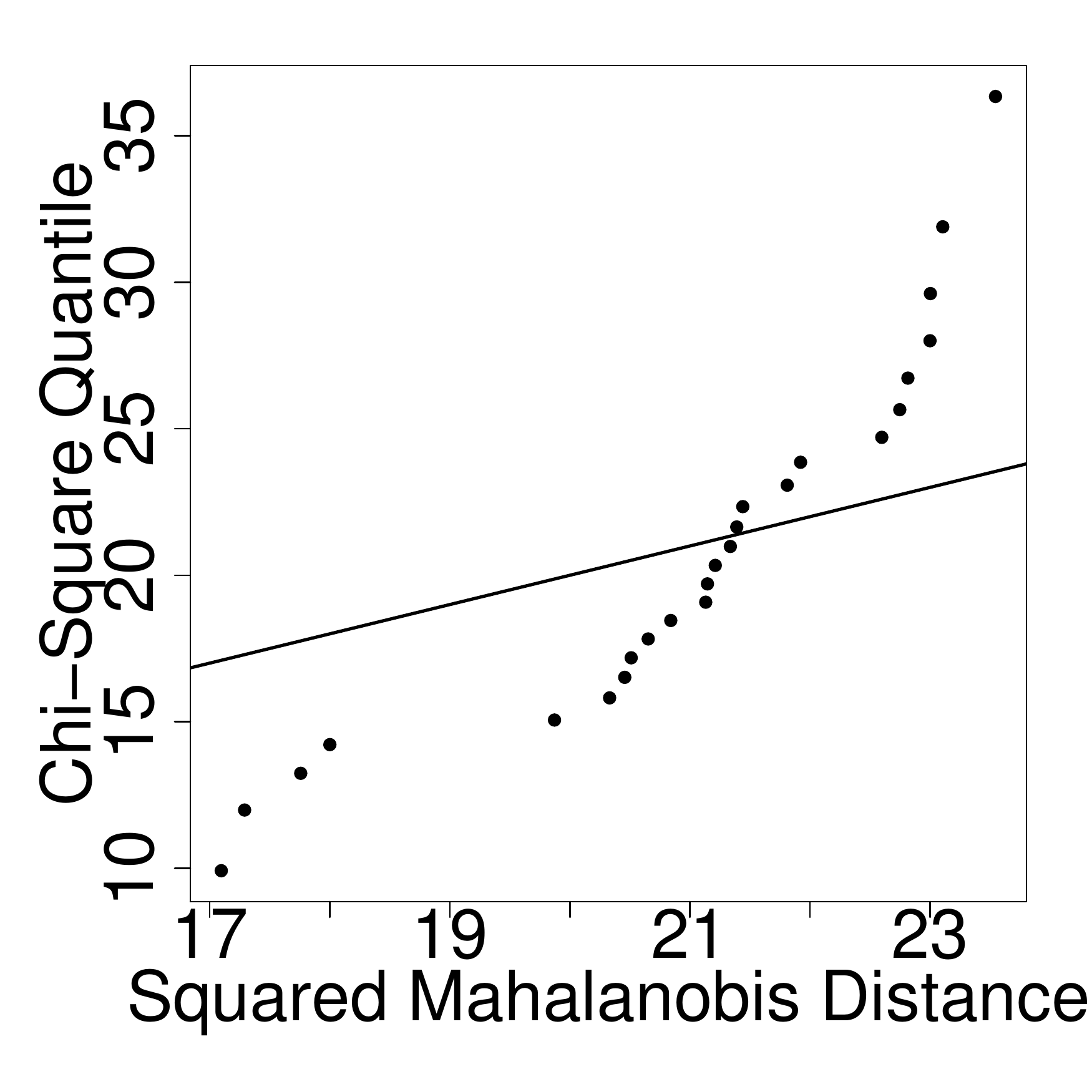}
    \label{fig:QG_qqplot_forecast_HMC_cycle_80}
  }
  \subfigure[HMC; t=1200]{
    \includegraphics[width=0.22\linewidth]{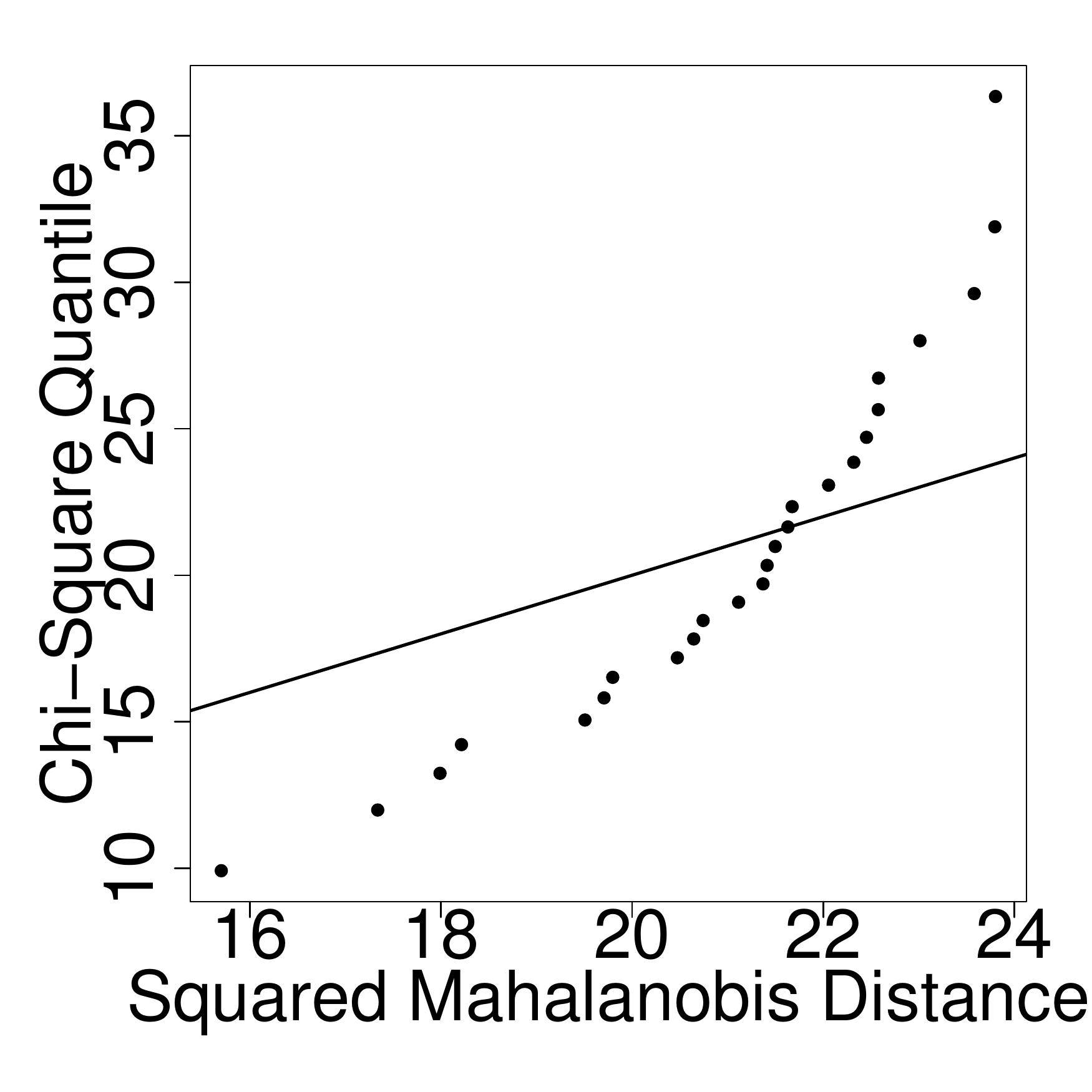}
    \label{fig:QG_qqplot_forecast_HMC_cycle_8096}
  }
  \\
  \subfigure[MC-\ClHMC; t=300]{
    \includegraphics[width=0.22\linewidth]{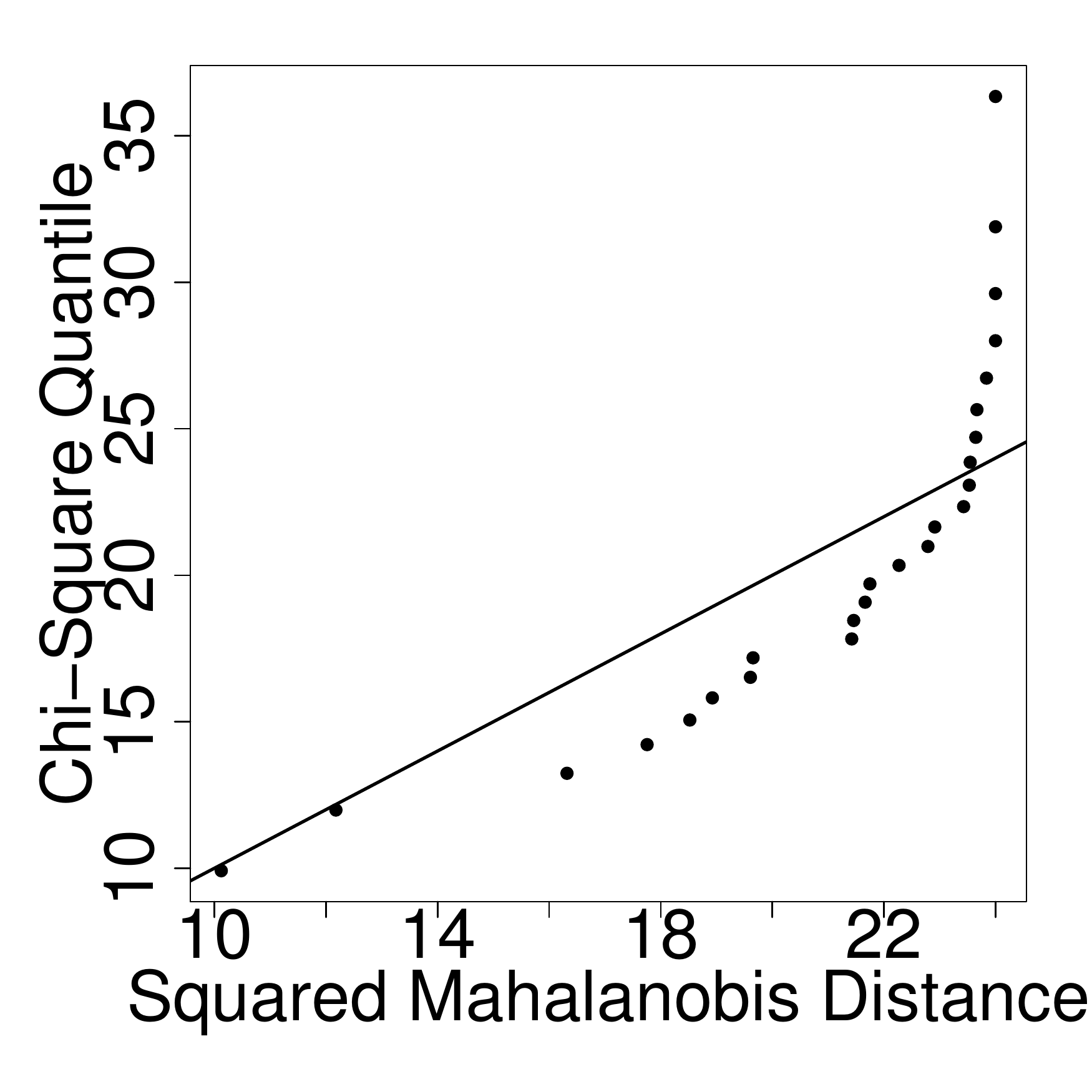}
    \label{fig:QG_qqplot_forecast_MC_ClHMC_AIC_cycle_23}
  }
  \subfigure[MC-\ClHMC; t=775]{
    \includegraphics[width=0.22\linewidth]{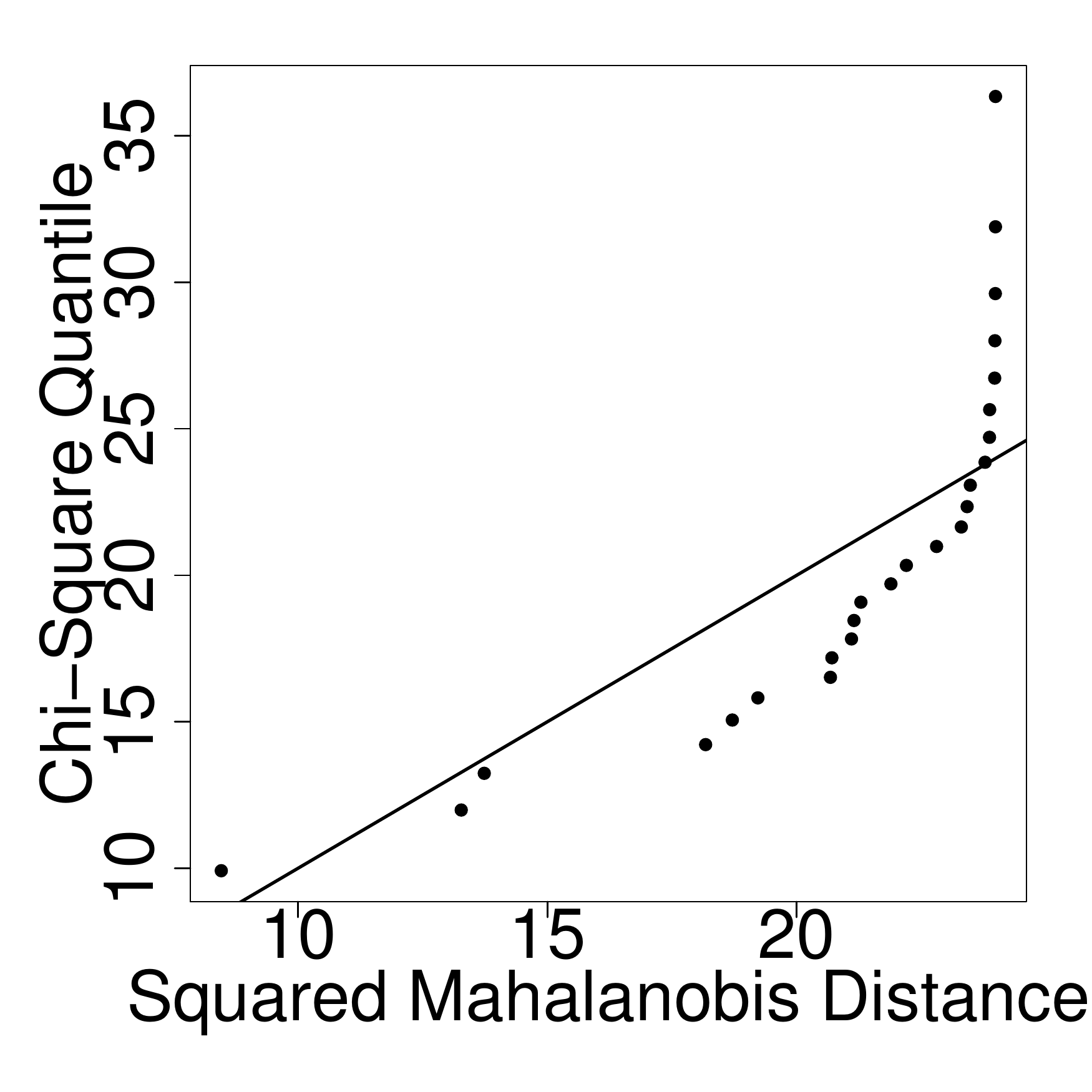}
    \label{fig:QG_qqplot_forecast_MC_ClHMC_AIC_cycle_61}
  }
  \subfigure[MC-\ClHMC; t=1000]{
    \includegraphics[width=0.22\linewidth]{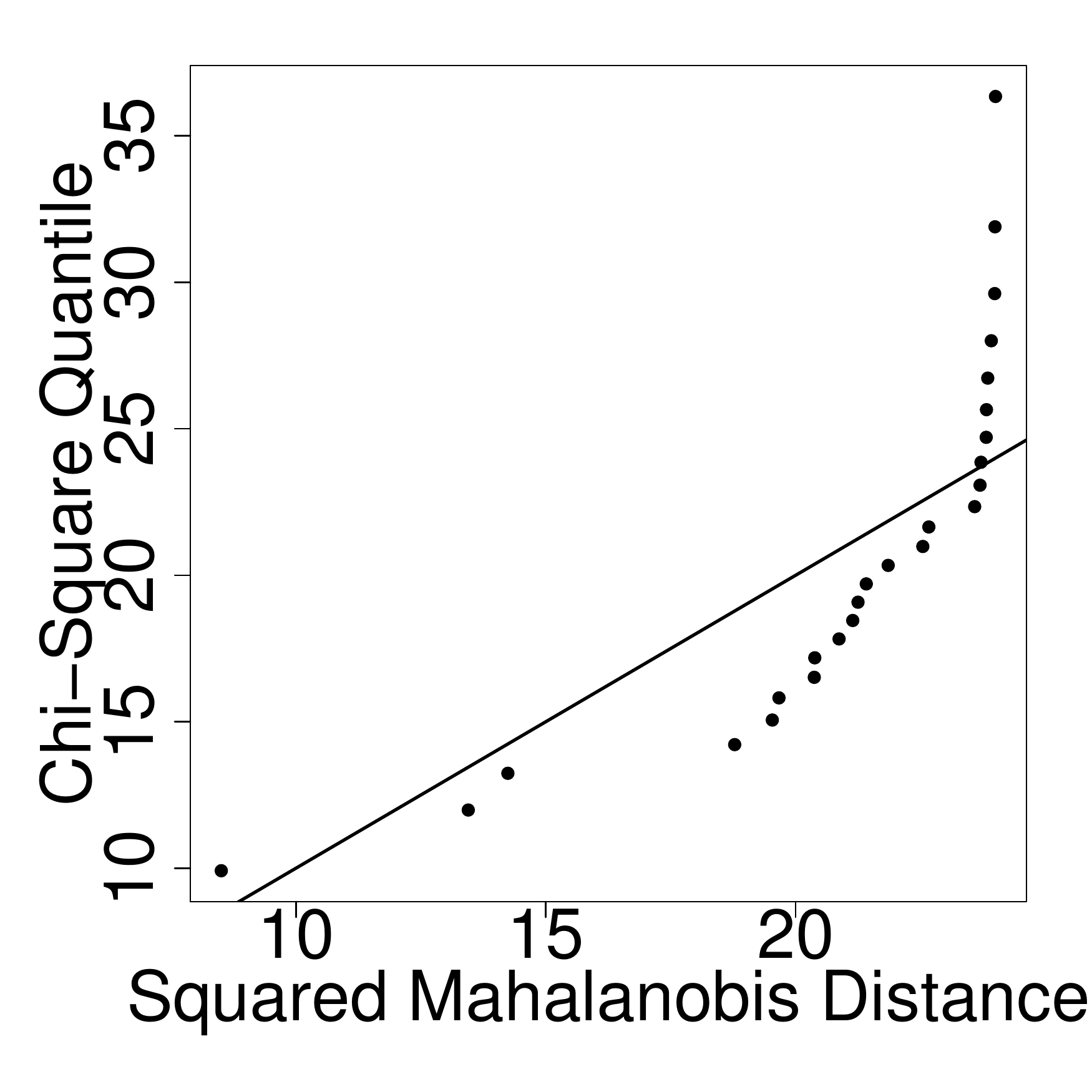}
    \label{fig:QG_qqplot_forecast_MC_ClHMC_AIC_cycle_80}
  }
  \subfigure[MC-\ClHMC; t=1200]{
    \includegraphics[width=0.22\linewidth]{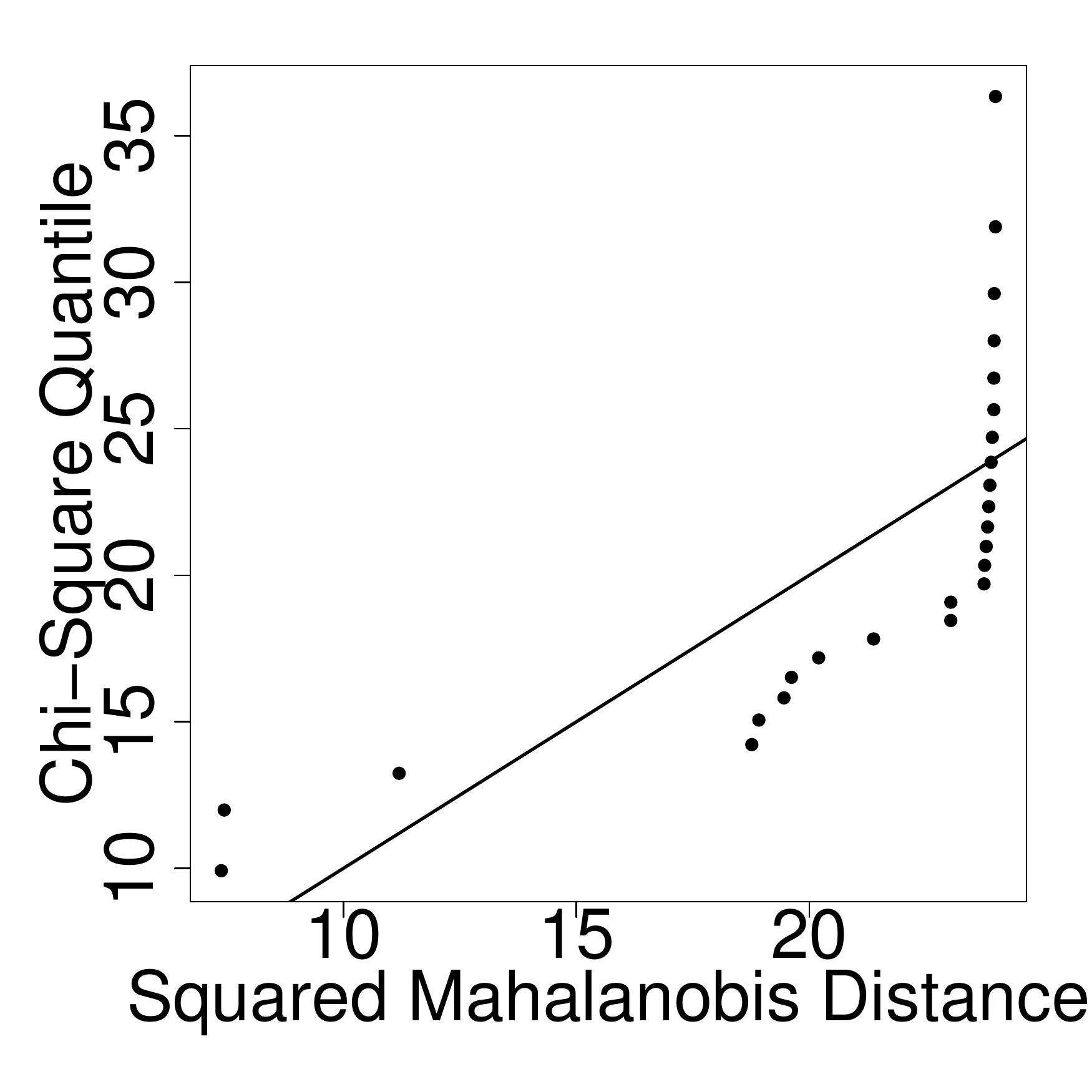}
    \label{fig:QG_qqplot_forecast_MC_ClHMC_AIC_cycle_8096}
  }
  \caption{Data assimilation with a linear observation operator. Chi-square Q-Q plots for the forecast ensembles obtained from propagating analyses of EnKF, HMC, and MC-\ClHMC filtering systems to times t=300, 775, 1000, and 1200 provide a strong indication of non-Gaussianity. The filtering methodology, and the assimilation time are given under each panel.  
           Localization is applied to the ensemble covariance matrix to avoid singularity while evaluating the Mahalanobis distances of the ensemble members.
  % A localized covariance matrix is used to calculate the Mahalanobis distances for the Q-Q plots.  %\sandu{using one covariance for all members, i.e., Gaussian-like setting? explain}
  }
  \label{fig:QG_linearH_Forecast_QQ_plots}
\end{figure}

To help decide whether to apply the original formulation of the HMC filter, or the proposed methodology, one can run  tests of non-Gaussianity on the forecast ensemble. To asses non-Gaussianity of the forecast several numeric or visualization normality tests are available, e.g., the Mardia test~\cite{mardia1970measures} based on multivariate extensions of skewness and kurtosis measures. Indication of non-Gaussianity can be found by visually inspecting several bivariate contour plots of the joint distribution of selected components in the state vector.
Visualization methods for multivariate normality assessment such as chi-square QQ-plots can be very useful as well.
Figure~\ref{fig:QG_linearH_Forecast_QQ_plots} shows several chi-square QQ-plots of the forecast ensembles generated from the result of EnKF, HMC, and MC-\ClHMC filters at different time instances. These plots show strong signs of non-Gaussianity in the forecast ensemble, and suggest that the Gaussian-prior assumption may in general lead to inaccurate conclusions.
%
%

%
%~~~~~~~~~~~~~~~~~~~~~~~~~~~~~~~~~~~~~~~~~~~~~~~~~~~~~~~~~~~~~~~~~~~
\subsubsection{Results with nonlinear wind-magnitude observations} 
% \label{Subsec:QG_Results}
%~~~~~~~~~~~~~~~~~~~~~~~~~~~~~~~~~~~~~~~~~~~~~~~~~~~~~~~~~~~~~~~~~~~
%
In the presence of a nonlinear observation operator the distribution is expected to show even stronger signs of non-Gaussianity. 
With stronger non-Gaussianity, the cluster methodology is expected to outperform the original formulation of the HMC sampling filter.

\begin{figure}[!htbp]
  \center
  \includegraphics[width=0.75\linewidth]{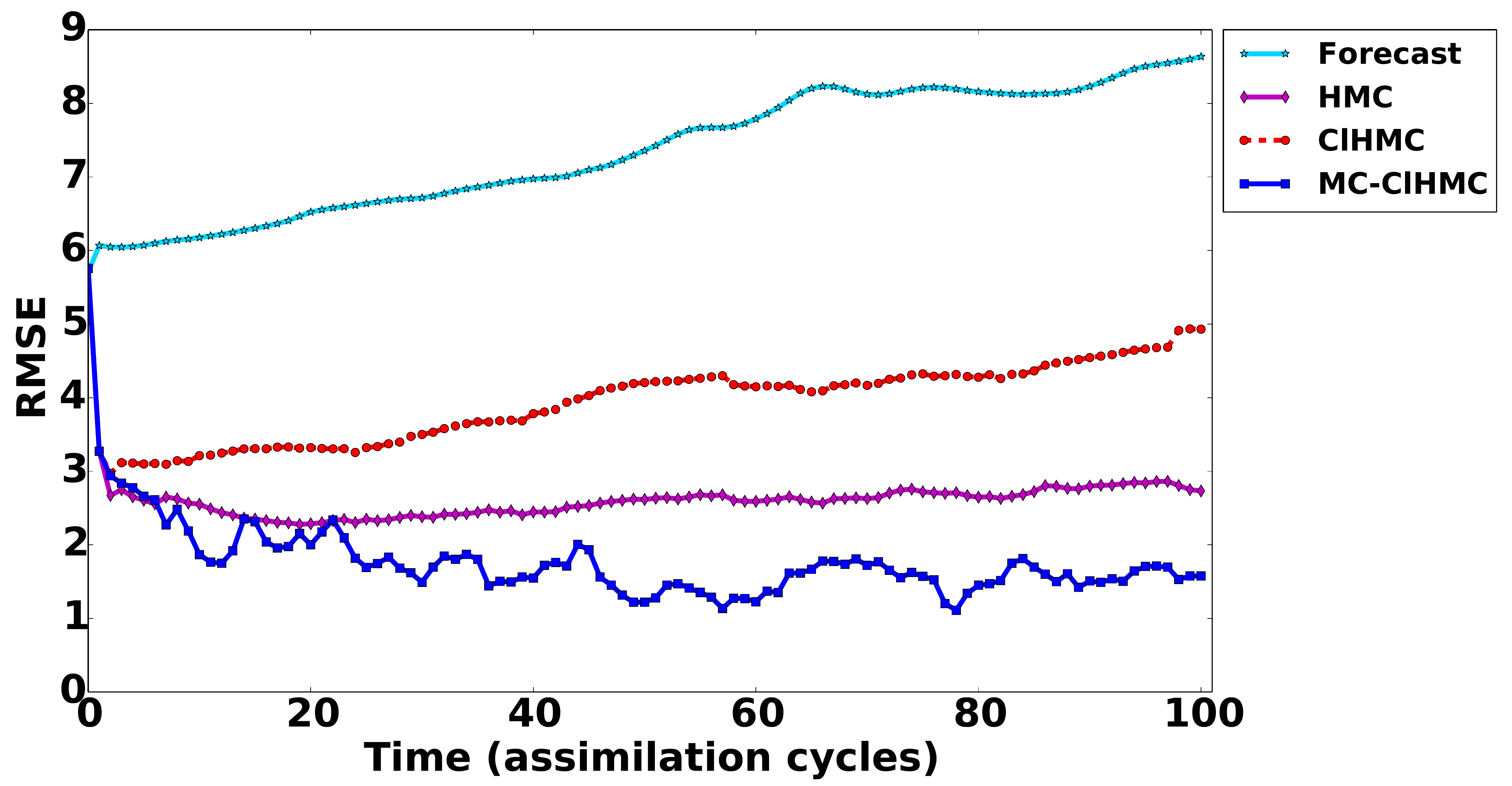}
  \caption{Data assimilation results with the nonlinear observation operator~\eqref{eqn:qg_wind_velocity}.   
   RMSE of the analyses obtained by HMC, \ClHMC, and MC-\ClHMC filtering schemes. 
   In this experiment, EnKF analysis diverged after the third cycle, and it's RMSE results have been omitted for clarity.
   }
  \label{fig:QG_NonLinearH_HMC_RMSE}
\end{figure}

Figure~\ref{fig:QG_NonLinearH_HMC_RMSE} shows RMSE results, with the nonlinear observation operator~\eqref{eqn:qg_wind_velocity}, for the analyses obtained by HMC, \ClHMC, MC-\ClHMC filtering systems. 
While EnKF diverges under the current settings after the third cycle (results omitted for clarity), HMC, \ClHMC, and MC-\ClHMC continue to behave similar 
to the case where the linear observation operator is used. 

\begin{figure}[!htbp]
  \center
  \subfigure[HMC]{
    \includegraphics[width=0.45\linewidth]{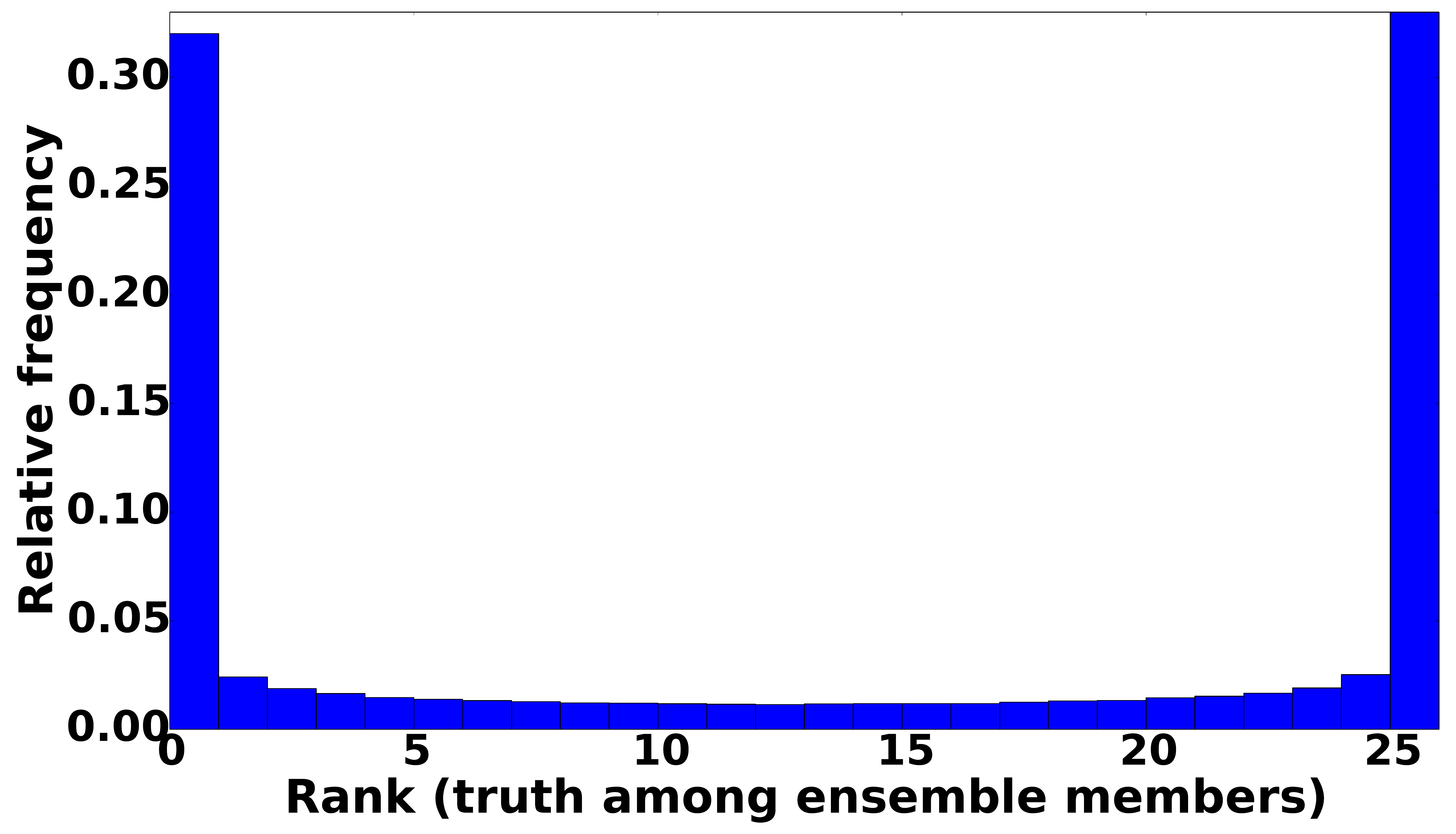}
    \label{fig:QG_NonLinearH_HMC_RankHist}
  }
  \hfill  
  \subfigure[\ClHMC+AIC]{
    \includegraphics[width=0.45\linewidth]{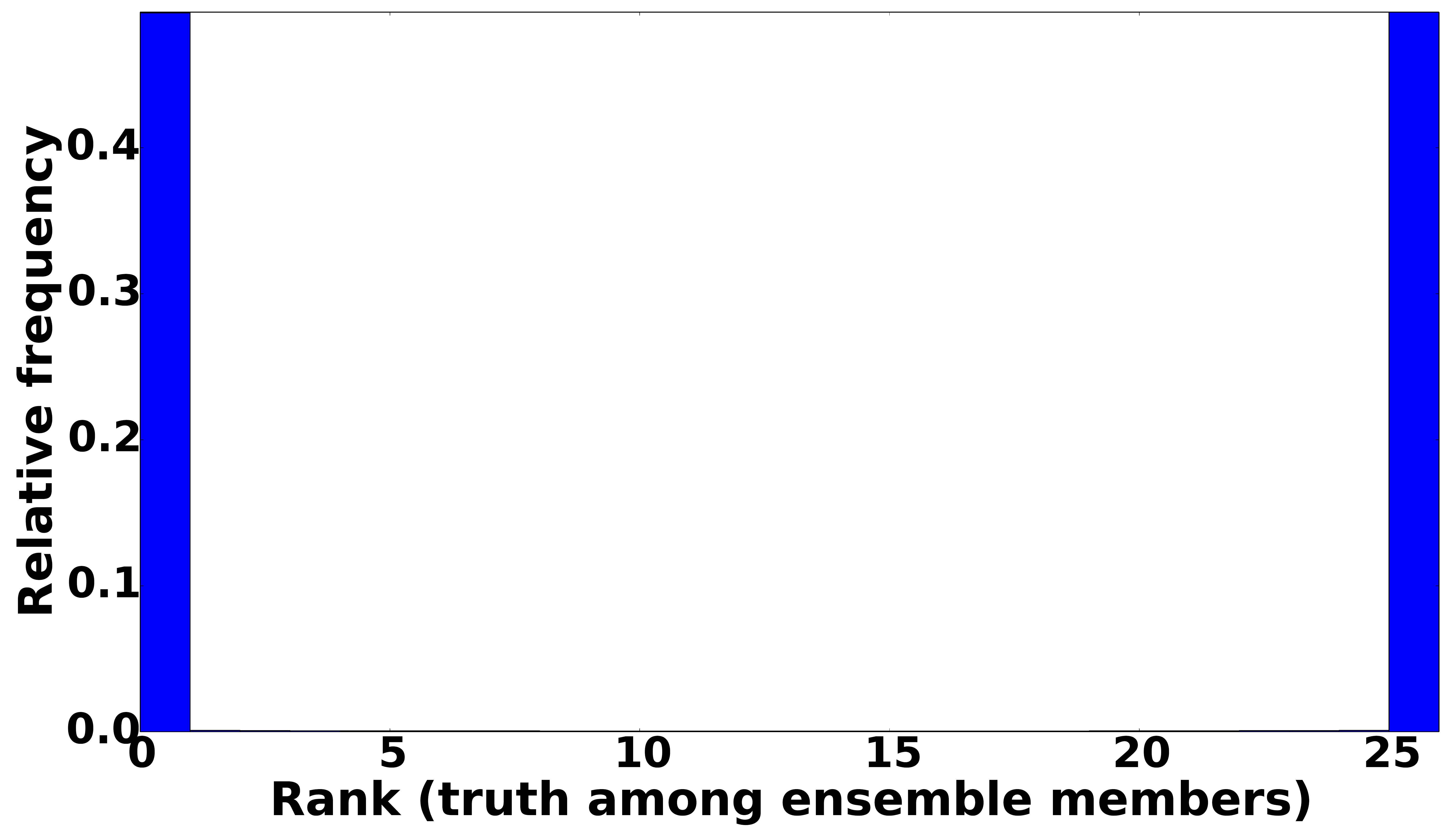}
    \label{fig:QG_NonLinearH_ClHMC_AIC_RankHist}
  }
  \\ % \quad % \hfill  
  \subfigure[MC-\ClHMC+AIC]{
    \includegraphics[width=0.45\linewidth]{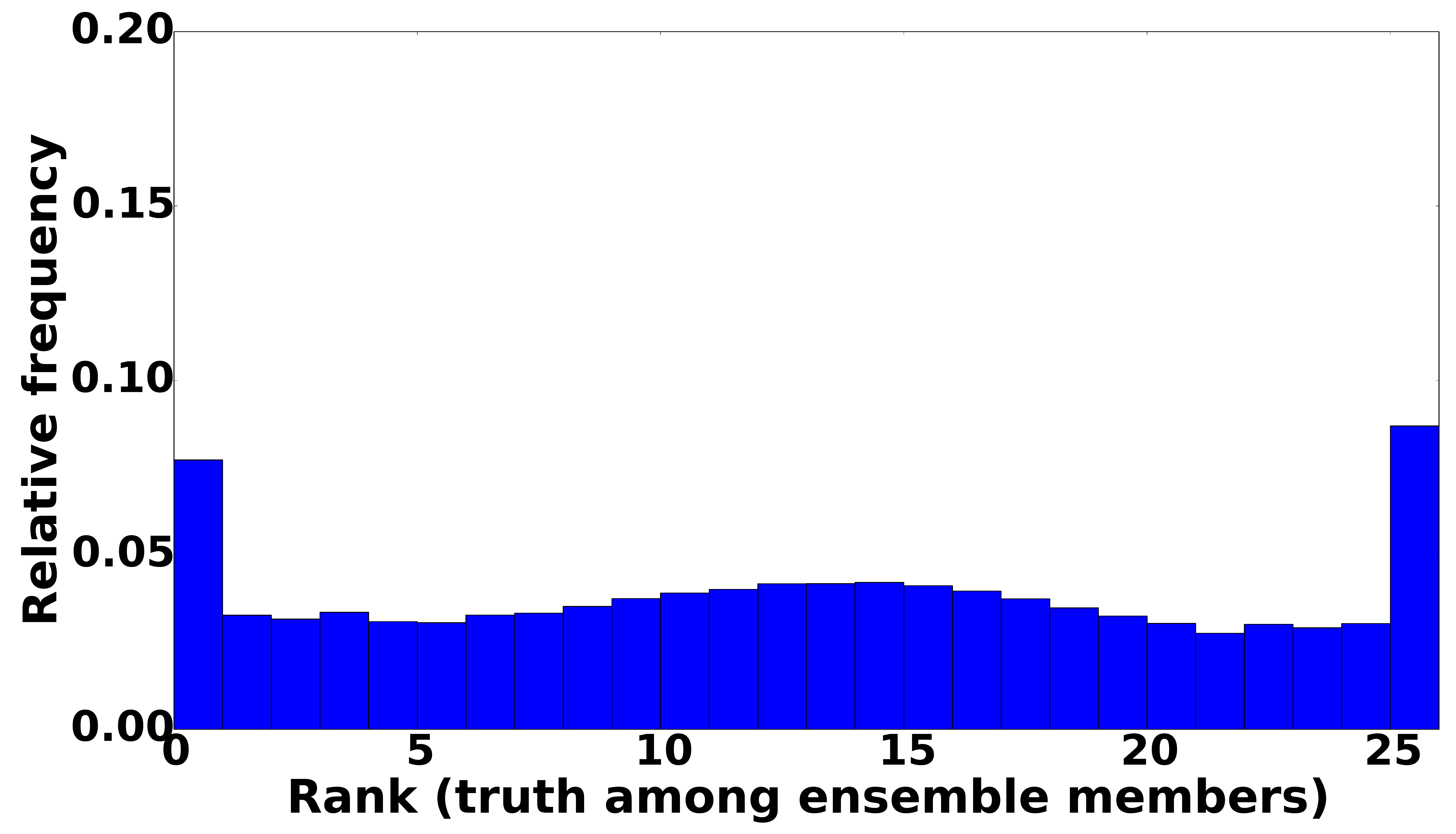}
    \label{fig:QG_NonLinearH_MC_ClHMC_AIC_RankHist}
  }
  \caption{Data assimilation results using the nonlinear observation operator~\eqref{eqn:qg_wind_velocity}.
	   The rank histograms of where the truth ranks among posterior ensemble members. 
	   The ranks are evaluated for every $16^{th}$ variable in the state vector (past the correlation bound) at $100$ assimilation times.
	   The filtering scheme used is indicated under each panel.
	   }
  \label{fig:QG_Nonlinear_H_RankHist}
\end{figure}

Figure~\ref{fig:QG_Nonlinear_H_RankHist} shows rank histograms of HMC, \ClHMC, and MC-\ClHMC, with a nonlinear observation operator. 
We can see that \ClHMC performance is similar to the case when the linear observation operator is used. 
It seems to be entrapped into a local minimum losing its dispersion quickly. 
The results of the MC-\ClHMC filter avoid this effect and show a reasonable spread.

The results presented here suggest that the cluster formulation of the HMC sampling filter is advantageous, 
especially in the presence of highly nonlinear observation operator, or strong indication of non-Gaussianity.

%____________________________________________________________________________
  \section{Conclusions and Future Work} \label{Sec:Conclusions}
  %____________________________________________________________________________
    %    
This work presents a new formulation of the HMC sampling filter for non-Gaussian data assimilation. The new formulation, named the Cluster HMC sampling filter (\ClHMC), relaxes the Gaussian prior assumption. 
The prior density is represented more accurately via a GMM fitted to the forecast ensemble. The initial formulation of the \ClHMC filter presented here is not expected to outperform the original HMC filter unless the sampler is capable of efficiently sampling multimodal distributions with modes separated by large regions of low probability. 
A multi-chain version of \ClHMC, namely MC-\ClHMC, is developed in order to achieve this goal. 

Numerical experiments are carried out using a nonlinear 1.5-layer reduced-gravity quasi geostrophic model in the presence of observation operators of different levels of nonlinearity.
The results show that the new methodologies are much more efficient that the original HMC sampling filter especially in the presence of a highly nonlinear observation operator.
    
The MC-\ClHMC is an algorithm that deserves further investigation. For example the local sample sizes here are selected based on the prior weight multiplied by the likelihood of the corresponding component mean. An optimal selection of the local ensemble size is required to guarantee efficient sampling from the target distribution.
    
Instead of using MC-\ClHMC filter, one can use \ClHMC with geometrically tempered Hamiltonian sampler as recently proposed in~\cite{nishimura2016geometrically}, such as to guarantee navigation between separate modes of the posterior. Alternatively, the posterior distribution can be split into $\nc$ target distributions with different potential energy functions and associated gradients. This is equivalent to running independent HMC sampling filters in different regions of the state space under the target posterior.
    
The authors have started to investigate the ideas discussed here, in addition to testing the proposed methodologies with automatically tuned HMC samplers.

\section*{Acknowledgments}
This work was supported in part  by the Air Force Office of Scientific Research (AFOSR) Dynamic Data Driven Application Systems program, by the National Science Foundation award NSF CCF (Algorithmic foundations) -- 1218454, and by the Computational Science Laboratory (CSL) in the Department of Computer Science at Virginia Tech.

\newpage
\bibliographystyle{plain}

  %____________________________________________________________________________
\section*{References}
  %____________________________________________________________________________
\bibliography{Bib/data_assim_general,Bib/data_assim_HMC,Bib/data_assim_kalman}

%=======================================================================================================================
\end{document}